\documentclass[12pt]{article}
\usepackage{amsmath,xcolor}
\usepackage{graphicx,psfrag,epsf}
\usepackage{enumerate,subcaption}
\usepackage{natbib,multirow}
\usepackage{url} 
\usepackage{algorithm, algpseudocode}
\usepackage{pdfpages}

\newcommand{\blind}{1}

\newcommand{\bfX}{{\bf X}}
\newcommand{\bfbeta}{\mbox{\boldmath $\beta$}}

\begin{document}

\def\spacingset#1{\renewcommand{\baselinestretch}%
{#1}\small\normalsize} \spacingset{1}


\if1\blind
{
		
  \title{\bf A Bayesian Multiscale Integrated Abundance Model for Estimating Latent Opioid Misuse Prevalence from Spatially Misaligned Data}
 \author{ Staci A. Hepler$^{1}$\thanks{Research reported in this publication was supported by the National Institute On Drug Abuse of the National Institutes of Health under Award Number R01DA052214. The content is solely the responsibility of the authors and does not necessarily represent the official views of the National Institutes of Health.} \and Brian N. White$^{2}$ \and Magdalena Cerd\'a$^3$ \and William C. Miller$^4$ \and Lance A. Waller$^5$ \and David M. Kline$^{2}$ }
   \date{$^{1}$Department of Statistical Sciences, Wake Forest University. Email: heplersa@wfu.edu \\
   $^2$ Department of Biostatistics and Data Science, Wake Forest University School of Medicine \\
   $^3$ Department of Chronic Disease Epidemiology, Yale School of Public Health \\
   $^4$ Department of Epidemiology, Gillings School of Public Health, University of North Carolina at Chapel Hill \\
   $^5$ Department of Biostatistics and Bioinformatics, Rollins School of Public Health, Emory University}
  \maketitle
} \fi

\if0\blind
{
  \bigskip
  \bigskip
  \bigskip
  \begin{center}
   {\LARGE\bf A Bayesian Multiscale Integrated Abundance Model for Estimating Latent Opioid Misuse Prevalence from Spatially Misaligned Data}
\end{center}
  \medskip
} \fi

\bigskip
\begin{abstract}
Estimating small area prevalence of opioid misuse is critical for targeting public health interventions, yet direct measures are unavailable and related surveillance data are often reported on misaligned areal units. We propose a Bayesian Multiscale Integrated Abundance (MIA) model for estimating latent opioid misuse prevalence by jointly analyzing multiple indirect surveillance indicators observed on different geographic supports. The model extends existing integrated abundance models by representing all source geographies through a common set of atomic spatial units formed by the intersections of observed areal supports. Latent prevalence at the atomic level is modeled using a Fisher noncentral hypergeometric distribution, which preserves county-level prevalence totals while respecting local population constraints. To enable scalable inference, we develop a two-stage compositional Markov chain Monte Carlo algorithm that combines customized sampling strategies and parallel computing. Simulation studies show the proposed approach reduces bias and root mean squared error relative to common downscaling methods. We apply the model to Ohio data from 2010-2023, integrating state-level survey estimates, county-level counts of opioid overdose deaths and treatment admissions, and ZIP code tabulation area-level counts of emergency medical services naloxone administrations. Results reveal substantial within-county heterogeneity and identify localized areas of elevated opioid misuse prevalence that would be missed by county-level analyses.
\end{abstract}

\noindent%

\spacingset{1.45} 

\section{Introduction}
\label{sec:intro}

The opioid epidemic remains a major public health crisis in the United States. Approximately 76\% of the 105,000 drug overdose deaths nationwide in 2023 involved opioids. While the total number of overdose deaths involving an opioid declined 4\% from 2022 to 2023, the number is still nearly ten times higher than it was in 1999 \citep{CDCopioid2025}. The impact of the opioid epidemic has been especially severe in the state of Ohio. In 2017, Ohio had the second highest age-adjusted opioid overdose death rate at 39.2 deaths per 100,000, and it reached a peak age-adjusted rate of 40.1 in 2021 before beginning to decline. In 2024, the age-adjusted opioid overdose death rate in Ohio had dropped to 20.1 deaths per 100,000 population. Despite being at its lowest level since 2014, that number remains over 13 times higher than the corresponding rate in 1999 \citep{KFFopioiddeaths}.

Effective public health responses to the opioid epidemic require information on where opioid misuse is occurring and how it varies across communities. While surveillance data such as opioid overdose death counts  provide proxy information for opioid misuse prevalence (OMP), any  surveillance indicator only provides partial information about the underlying population at risk since not all people who misuse opioids (PWMO) experience an observable outcome. Ideally, estimates of OMP would be used to evaluate policy, guide resource allocation, identify health service gaps, and highlight emerging hotspots. However, OMP is not observed directly at finer spatial scales. The National Survey on Drug Use and Health (NSDUH) generates estimates of OMP, but these are at the state and multi-year level \citep{NSDUH}. Thus, small area estimates of OMP must be inferred from existing indirect data sources such as overdose death counts, counts of treatment admissions, and counts of naloxone administrations.

Indirect estimation is a common strategy for estimating hidden or hard-to-reach population sizes and has been applied to estimate OMP and opioid use disorder. Capture-recapture \citep{Jones2016,Barocas2018,Min2020,thompson_prevalence_2023, wang_intersectional_2024}, benchmark-multiplier \citep{keyes_what_2022,wang_estimation_2024}, and multi-parameter estimation of prevalence \citep{markoulidakis_prevalence_2024,markoulidakis_multi-parameter_2026} models have all been used to leverage multiple sources of data to estimate population prevalence. Each require at least some data that are individually-linked across data sources or estimates derived from surveys and studies with individual-level data. However, government agencies in many jurisdictions only release data in aggregate form to protect privacy and comply with legal obligations. These restrictions do not allow for use of methods based on individual-level identification and linkage and require alternative approaches based exclusively on aggregate data sources. 

Given privacy and aggregation restrictions, methods have been developed that rely exclusively on aggregate data. One such method uses a top-down approach with a multinomial distribution to infer county-level counts of opioid use disorder across the United States using county-level covariates to downscale state-level counts \citep{peterson_bayesian_2025}. Another alternative is the Bayesian Integrated Abundance (BIA) model, which was developed by \citet{hepler_integrated_2023} to infer latent county-level OMP by jointly modeling multiple county-level surveillance outcomes. By modeling multiple aggregate outcomes jointly with state-level NSDUH survey estimates of OMP, the BIA overcomes identifiability issues known to occur with abundance models. They applied the BIA to data from Ohio on opioid overdose deaths and treatment admissions to infer yearly county-level OMP. Another advantage of the BIA is it permits quantification of outcome-specific rates among PWMO, which is often more meaningful than outcome rates per population. The BIA has since been applied to other states including New York \citep{santaella2024estimation} and North Carolina \citep{kline2025estimating}, and it has been used to identify counties where there are discrepancies between OMP and buprenorphine prescribing \citep{mcknight2023descriptive}.

The approaches described above generate estimates at the county-level. Many local health departments are organized at the county level so county-level OMP estimates are particularly valuable for state health officials tasked with allocating resources.  However, counties are large, heterogeneous regions, and county-level estimates obscure within-county variability and can thus fail to identify higher resolution geographical regions with elevated OMP, which is critical information for local health departments. As one example, \cite{mcmaster2025drug} examined drug-related mortality at the Census tract level within Cuyahoga County, Ohio. They found that while the county average drug-related mortality rate is near the statewide average, there are Census tracts within the county whose rates are substantially higher. The authors emphasize the need for more granular geographic analysis to facilitate targeted public health interventions. While this analysis was focused on drug-related mortality, we expect there is similarly large within-county variation in OMP. 

While most surveillance data on opioid-related indicators are available at the county level, there are instances where data on a subset of indicators are available at a finer spatial resolution such as Census tracts or ZIP codes. A major challenge when integrating such data sources is that the spatial units on which they are reported are often misaligned and sometimes even non-nested, creating an areal spatial change-of-support (COS) problem. 

\cite{Gotway2002} provide a broad overview to combining misaligned data. \cite{bradley2016bayesian} and \cite{Bradley2015} used a Bayesian COS ``bottom up'' strategy to model misaligned data assuming observed data on areal units were aggregates of a spatial point process. While this has the advantage of allowing the user to integrate to any desired areal spatial support, this was designed for settings where one is modeling misaligned data on the same quantity of interest. ``Top-down'' approaches have commonly been used for COS with areal count data. These use a hierarchical Poisson regression model to model a latent spatial process on the finest areal spatial resolution that is linked to the outcomes through aggregation to the coarser geographical scales. This framework has been used to study the relationship between a response variable and misaligned explanatory variables and to downscale the coarse response variable to a finer spatial resolution. See, for example, \cite{Mugglin1998, Mugglin1999, Mugglin2000} and \citet{peterson_bayesian_2025}. Although there is a deep literature on spatial COS, the emphasis is on downscaling an observed variable. In our setting, we do not observe county- or sub-county-level OMP. Instead, methods to indirectly infer sub-county latent OMP that formally integrate multiple sources of information across spatially misaligned areal units are needed. 


In this paper, we propose the multiscale integrated abundance (MIA) model that extends the Bayesian integrated abundance framework of \cite{hepler_integrated_2023}. The proposed model jointly analyzes surveillance indicators observed on multiple geographic supports and estimates OMP at both county and sub-county resolutions. By representing all source geographies through a common set of atomic spatial units formed by their intersections, the model accommodates spatially misaligned data while preserving consistency across scales. The resulting framework allows public health agencies to estimate latent OMP at a much finer geographic resolution than is possible using county-level surveillance data alone. 


There are two main contributions in this paper. The first is the development of the MIA model to provide multiscale estimates that can be used to address the needs of both state and local public health officials. A novel aspect of the MIA is the integration of the Fisher Noncentral Hypergeometric (FNCH) distribution as a model for the latent atomic-level process to enforce sub-county constraints related to small area population. This model also provides estimates of OMP at all geographical levels that have standard epidemiological interpretations that can be readily used by health department staff and help to ease integration into practice. The second contribution is the development of a two-stage Markov chain Monte Carlo (MCMC) approach that novelly combines parallel computing with strategic sampling algorithms to efficiently satisfy model constraints while remaining scalable and generalizable. While our work is motivated by the need for estimates of OMP in Ohio, the MIA model can more generally be applied to prevalence estimation of hard-to-reach populations where only data on aggregate indirect indicators are available. 

Section \ref{sec:data} introduces the data from Ohio that motivated our methodological development. The MIA model and two-stage MCMC approach are described in Section \ref{sec:method}. In Section \ref{sec:sim}, we present simulation results that compare the MIA model's sub-county estimates to estimates that would be obtained through commonly used downscaling approaches, and in Section \ref{sec:app}, we apply the MIA to estimate yearly, sub-county OMP in Ohio.

    

\section{Data and Motivating Application}
\label{sec:data}

Our proposed model integrates multiple data sources at varying spatial and temporal scales, with the goal of performing multiscale inference on OMP in Ohio. Below, we describe the different data sources used, broken down by spatial resolution.

\subsection{State-Level Data}

The only direct data source available on OMP comes from the National Survey on Drug Use and Health (NSDUH) \citep{SAMHSAsurvey}. Prior to 2015, this survey provided estimates of past-year nonmedical opioid use, and post-2015 it estimated past-year opioid misuse. Despite the change in survey language, we treat these measures as comparable over time. The NSDUH reports multi-year averaged estimates of statewide prevalence for Ohio, along with associated standard errors. We use estimates spanning 2007–2022 to inform the statewide prevalence trajectory. The survey estimates and standard errors are shown in Supplemental Material (SM) Table 1. 


\subsection{County-Level Data}

The county-level data included is the same as in \cite{hepler_integrated_2023}, but updated to include more recent years. More specifically, for each of Ohio's 88 counties and each year from 2010 - 2023, we use two surveillance indicators: opioid overdose death counts and counts of treatment admissions. Death counts come from the Ohio Public Data Warehouse and include yearly county-level counts of deaths where the death certificate mentions poisoning from any opioid under the ICD-10 codes T40.0-T40.4 and T40.6. Deaths are indexed by the county of residence of the decedent. The treatment admissions counts represent unique patients and come from the Ohio Department of Behavioral Health through a data use agreement that includes counts of admissions to residential, outpatient, or intensive outpatient treatment for opioid use disorder. Treatment counts under 50 are suppressed. County-level population data is from the National Center for Health Statistics and obtained through the Ohio Public Health Data Warehouse \citep{OPHDW}. Maps of the county-level indicators are shown in SM Figures 1 and 2.

Our county-level model includes other county-level variables as covariates. Sociodemographic variables come from the American Community Survey (ACS) 5-year estimates. The ACS variables included are the poverty rate, unemployment rate, proportion with at least a high school degree, and the proportion on food stamps. We also include data on opioid prescribing rates, which is publicly available from the Ohio Automated RX Reporting System. Additionally, we compile county-level indicators of being designated a health professional shortage area (HPSA) or a medically underserved (MU) area from the Health Resources and Services Administration, a high impact drug trafficking area (HIDTA) from the Drug Enforcement Agency, a metropolitan statistical area (MSA) from the United States Census Bureau, and an indicator for having an interstate highway intersect the county. 

\subsection{ZCTA-Level Data}

Naloxone is an opioid antagonist used to reverse overdose. For 2017 to 2023, we have indicator data on emergency medical services (EMS)-administered naloxone counts aggregated to the ZIP-code level. These data were obtained from the Ohio Department of Public Safety Division of Emergency Medical Services, through the EMS Incident Reporting System \citep{ohio_emergency_medical_services_nalxone_2026}. Counts reflect EMS administrations recorded by EMS staff regardless of the final cause of the incident and do not reflect unique patients. We process all ZIP-code level data to the ZIP-code tabulation area (ZCTA) level because U.S. Census data sources provide demographic information at the ZCTA-level. This processing is done using a crosswalk obtained from the Missouri Census Data Center. For a detailed description, see \cite{murphy2025integrating}. We also removed remaining ZCTA that corresponded to single delivery-point address. The final dataset includes 1063 ZCTAs. The ZCTA-level rates of EMS administered naloxone are shown in SM Figures 3-5.

ZCTA-level covariates are included to capture spatial variation in EMS-administered naloxone rates. We include the proportion of each ZCTA who are unemployed and the proportion with at least a Bachelor's degree from the ACS 2018 - 2022 5-year estimates. We construct indicators of urbanicity based on the rural-urban commuting area (RUCA) codes. We define ZCTAs with RUCA codes between 1-3 to be `metropolitan' and ZCTAs with RUCA codes between 4-6 to be `micropolitan,' and we include two indicator covariates for each designation with codes between 7-10 as a `rural' reference group. 


\subsection{Atom-Level Data}

\begin{figure}
    \centering
    \includegraphics[trim=4.5in .7cm 4.5in .65cm, clip, width=0.4\linewidth]{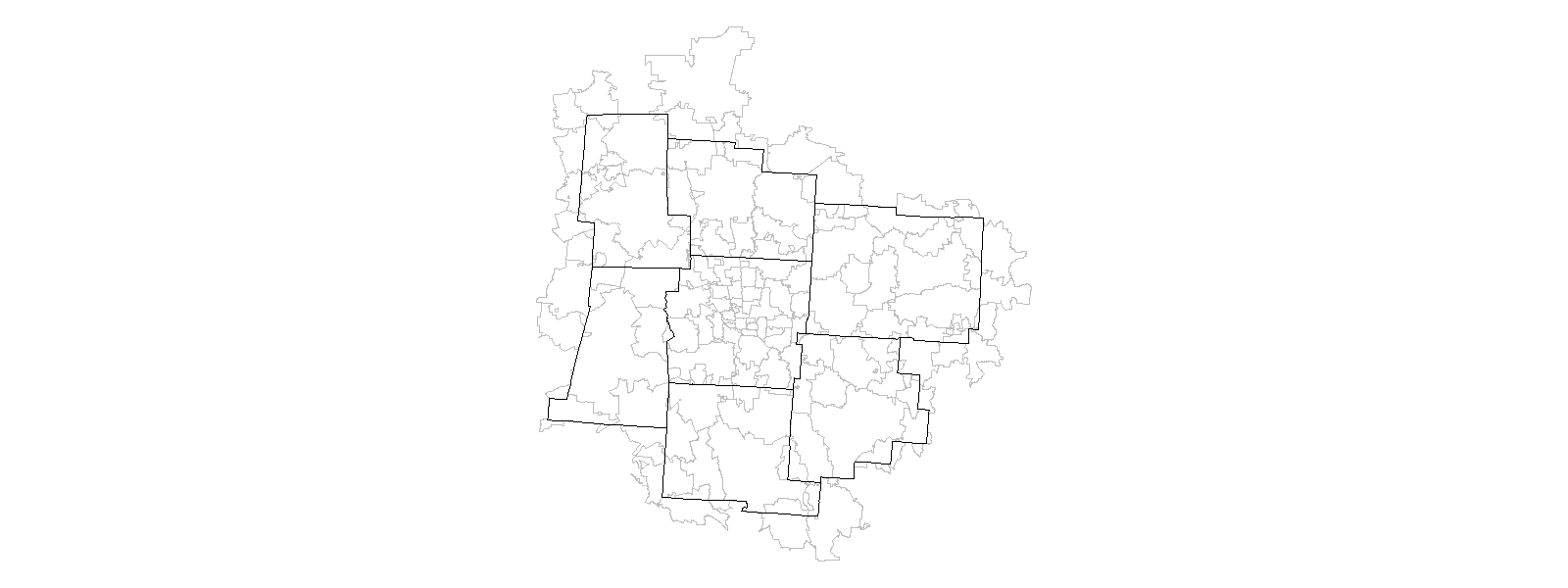}
        \includegraphics[trim=4.5in .7cm 4.5in .65cm, clip, width=0.4\linewidth]{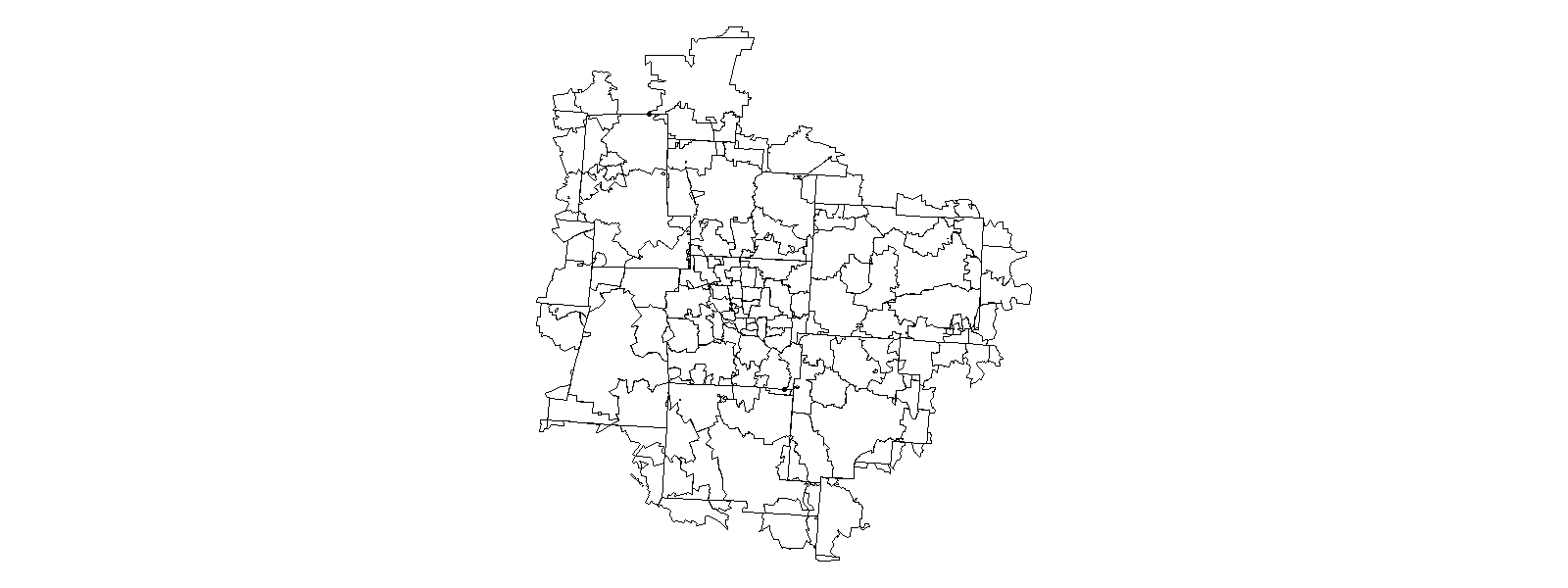}
    \caption{The left map shows the two source zones - Central Ohio counties (black lines) and the ZCTA that intersect them (gray lines). The right plot shows the resulting atoms (target zones) formed by intersecting the source geographies.}
    \label{fig:CentralOHatoms}
\end{figure}

Throughout, we will follow the terminology of \cite{Mugglin1999}, \cite{bradley2016bayesian}, and \cite{trevisani2013spatial} and define the source zones to be the areal units where count data are available, and the target zones as the area where we wish to make inference. Atoms are the spatial units constructed by intersecting our source geographies. In our case, counties and ZCTAs serve as the source support and atoms, which are the intersection of counties and ZCTAs, are the target support. Note that while counties and ZCTAs are non-nested, atoms are nested both within counties and ZCTAs. This construction ensures all modeled quantities can be expressed as sums over a common set of atomic units, permitting multiscale inference at both the target and source levels of geography. Our study consists of 1670 atoms across Ohio. Figure \ref{fig:CentralOHatoms} illustrates  source and target geographies in our application zoomed into central Ohio.


We obtain atom-level population counts using the 2022 Geographic Correspondence Engine from the Missouri Census Data Center \citep{mcdc_geocorr2022}. This provided county-level population counts from the 2020 Census and the proportion of the county that resided within each atom formed by intersecting counties and ZCTAs. We also included the proportion of the atom population that is unemployed and the proportion with at least a bachelor's degree. These atom-level covariates were constructed by assuming the ZCTA-level rates from the 2018-2022 5-year ACS estimates were constant for all atoms within the ZCTA. Maps of the atom-level covariates are included in SM Figures 6 and 7.

\section{Bayesian Multiscale Methodology}
\label{sec:method}

We begin by briefly reviewing the county-level BIA that serves as the foundation for the proposed multiscale framework. This model integrates multiple county-level surveillance data sources with state-level survey information to infer latent OMP at the county level. A full description of this model is in SM Section 2.1. In Section \ref{ss:model}, we describe the multiscale extension. Without loss of generality and for ease of exposition, we describe the model assuming count-valued aggregate data on counties and ZCTAs; however, the framework can be applied for any type of areal spatial units (e.g., Census tracts, cities, etc) and straightforwardly extended to more than two types of geographical units. 


\subsection{County-Level Bayesian Integrated Abundance Model}
\label{ss:BIA}

Let $Y_{it}^{(k)}$ denote the observed count for indicator $k=1,...,K$ at county $i=1,...,n^{(C)}$ during year $t=1,...,T$. In our application, $k=1$ corresponds to opioid overdose deaths and $k=2$ corresponds to treatment admissions. Conditional on the county-level latent number of PWMO, $N_{it}$, we assume
\begin{equation}
\label{eq:ctydata}
    Y_{it}^{(k)}|N_{it},p_{it}^{(k)} \stackrel{ind}{\sim} \text{Binomial}\left(N_{it},p_{it}^{(k)}\right),
\end{equation}
where $p_{it}^{(k)}$ represents the probability that an individual who misuses opioids experiences indicator $k$. We model the indicator probabilities using logistic regression
\begin{equation*}
\text{logit}\left(p_{it}^{(k)}\right)=\mu_{t}^{(k)}+\bfX_{it}^{(k)} \bfbeta^{(k)}+ f^{(k)}_{it}+\epsilon_{it}^{(k)},
\end{equation*}
where $\mu_{t}^{(k)}$ is a time-varying intercept capturing statewide temporal trends, $\bfX_{it}^{(k)}$ is a vector of centered covariates (excluding an intercept), and $\bfbeta^{(k)}$ is a vector of regression coefficients. To capture spatial and temporal dependence, $f^{(k)}_{it}$ is modeled with an intrinsic conditional autoregressive (ICAR) model with first-order temporal autocorrelation, and $\epsilon_{it}^{(k)}$ accounts for additional unexplained heterogeneity. Note that in our application, treatment admission counts less than 50 are censored. We account for this in the model using an interval censored likelihood as in \cite{Famoye2004}. Specific details can be found in SM Section 2.1.


A challenge in this setting is that surveillance data alone do not identify the overall level of OMP \citep{stoner_hierarchical_2019}. To address this, we incorporate data from statewide National Survey on Drug Use and Health (NSDUH) estimates, which provide estimates of statewide OMP over multi-year periods, along with standard errors of the estimates. Let $S_{a:b}$ denote the estimate of the statewide OMP between years $a$ and $b$. The estimates for our application are shown in SM Table 1. We apply a logit-transformation to the estimates and use a first-order Taylor expansion to approximate the standard error. 
Let $\mu_t$ denote the true latent statewide OMP in year $t$. We assume $\text{logit}(\mu_t) = \beta_0^{\mu} + \beta_1^{\mu} t$, and 
\begin{align}
    \text{logit}(S_{a:b})|\beta_0^{\mu},\beta_1^{\mu} \sim N\left( \frac{1}{b-a+1}\sum_{t=a}^b \text{logit}(\mu_t), \hat{se}_{a:b}^2 \right).
    \label{eq:surveymulti}
\end{align} 

The target of inference in the county-level model is the latent number of PWMO, $N_{it}$, and the prevalence ratio, $\lambda_{it}$.  Assume
   $ N_{it}|\lambda_{it} \stackrel{ind}{\sim} \text{Binomial}(P_{it}, \mu_t \lambda_{it}),$
where $P_{it}$ is the known population of county $i$ during year $t$ and $\mu_t$ is the statewide OMP that also appeared in \eqref{eq:surveymulti}. The prevalence ratio is relative to the statewide OMP and modeled as $ \log(\lambda_{it})= \mathbf{W}^{(C)}_{it} \boldsymbol{\gamma} + u_{it}+v_{it},$ where 
$\mathbf{W}^{(C)}_{t}$ is a design matrix for time $t$ containing centered covariates without an intercept,  $v_{it} \overset{iid}{\sim} N(0,\sigma^2_v)$, and $u_{it}$ is a spatio-temporal random effect defined similarly to $f_{it}^{(k)}$. See  SM Section 2.1 for full specification of prior distributions.



\subsection{Multiscale Integrated Abundance Model}
\label{ss:model}

The multiscale integrated abundance (MIA) model extends the model in Section \ref{ss:BIA} to include additional indicators at a finer spatial resolution. This top-down approach yields multiscale estimates of latent OMP while also quantifying covariate effects at the different resolutions.  
Note that all modeling components from Section \ref{ss:BIA} are still incorporated within the full hierarchical model and are as defined above and in SM Section 2.1. 


Let $A_{ij}$ denote atom $j$ within county $i$ where $i=1,...,n^{(C)}$ and $j=1,...,n^{(c)}_i$ such that $n^{(c)}_i$ is used to denote the number of atoms within county $i$. Similarly, let $B_{\ell k}$ denote atom $k$ within ZCTA $\ell$, where $\ell=1,...,n^{(Z)}$ and $k=1,...,n^{(z)}_{\ell}$ such that $n^{(Z)}$ denotes the total number of ZCTAs and $n^{(z)}_{\ell}$ the number of atoms within ZCTA $\ell$. Observe that $\cup_{i,j} A_{ij} = \cup_{\ell,k} B_{\ell k}$, but $A$ and $B$ will be used when we are describing an atom \textit{within} a county versus \textit{within} a ZCTA, respectively. Let $n^{(a)}$ denote the total number of atoms. 

\subsubsection{ZCTA-Level Data Model}
\label{ss:zctamod}

Let $Z_{\ell t}^{(o)}$ be the observed count for ZCTA-level indicator $o$ at ZCTA $\ell$ during year $t$.  Assume
$Z_{\ell t}^{(o)} = \sum_j Z_t^{(o)}(B_{\ell j}),$ where $Z_t^{(o)}(B_{\ell j})$ denotes the \textit{latent} count of indicator $o$ for atom $j$ within ZCTA $\ell$. For each atom, assume 
    $Z_t^{(o)}(B_{\ell j}) \mid N_t(B_{\ell j}), q_t^{(o)}(B_{\ell j}) \sim \text{Poisson}\left(N_t(B_{\ell j}) q_t^{(o)}(B_{\ell j})\right),$
where $N_t(B_{\ell j})$ is the latent atom-level count of PWMO in year $t$, and $q_t^{(o)}(B_{\ell j})$ is the rate at which PWMO are counted by indicator $o$ in that atom during that year. Marginalizing over the latent atom-level counts yields 
\begin{equation}
    Z_{\ell t}^{(o)} \mid \{N_t(B_{\ell j}), q_t^{(o)}(B_{\ell j}) : j=1,...,n_{\ell}^{(z)} \} \sim Poisson\left( \sum_{j=1}^{n_{\ell}^{(z)}} N_t(B_{\ell j}) q_t^{(o)}(B_{\ell j}) \right).
    \label{eq:naloxonemodel}
\end{equation}

We model the atom-level rates $q_t^{(o)}(B_{\ell j})$ with a log-linear model of the form
\begin{equation}
    \log(q_t^{(o)}(B_{\ell j})) = \bfX_t^{(o)}(B_{\ell j}) \bfbeta^{(o)} + f_t^{(o)}(B_{\ell j}),
    \label{eq:zctarate}
\end{equation}
where $\bfX_t^{(o)}(B_{\ell j})$ is a vector of atom-level covariates with regression coefficients $\bfbeta^{(o)}$, and $f_t^{(o)}(B_{\ell j})$ is a spatio-temporal random effect. Note that the spatial dimension of $f_t^{(o)}(\cdot)$ is large when $n^{(a)}$ is big. We reduce the spatial dimension by assuming $f_t^{(o)}(B_{\ell j}) = \mathbf{B}(B_{\ell j})'\boldsymbol{\omega}^{(o)}_t,$ where $\mathbf{B}(B_{\ell j})$ is a $q$-dimensional vector of Moran's I basis functions \citep{hughes2013dimension}. The coefficients in the $q$-dimensional vector $\boldsymbol{\omega}^{(o)}_t$ are modeled with a temporal autoregressive process. See SM Section 2.2 for the full specification of $f_t^{(o)}(B_{\ell j})$. 

\subsubsection{Atom-Level Model for Latent OMP}

We take a top-down approach that allocates county-level latent counts of PWMO to finer spatial units, which requires a model that simultaneously enforces two structural constraints. First, the atom-level counts must sum to the county total, $\sum_{j=1}^{n_i^{(c)}} N_t(A_{ij}) = N_{it}$,  and second, atom counts cannot exceed the atom population, $N_t(A_{ij}) \leq P_t(A_{ij})$. We achieve this by modeling the vector of atom-level counts within each county using the Fisher noncentral hypergeometric distribution (FNCH), which arises when a fixed number of individuals are allocated across categories with unequal selection probabilities and finite capacities \citep{harkness1965properties, fog2008sampling}. 
This preserves the county-level totals while allowing covariates and spatio-temporal random effects to influence the relative allocation across atoms. Compared to a multinomial distribution, which ignores population limits, or independent binomial models, which fail to enforce the sum constraint, the FNCH satisfies both constraints while  providing a desirable interpretation as it models atom-level OMP. More formally, if
$\mathbf{N}_t(A_{i\cdot})=\left\{N_t(A_{ij}) : j=1,...,n_i^{(c)}\right\}$ denotes the vector of atom-level counts within county $i$, then assuming the FNCH implies the probability function of $\mathbf{N}_t(A_{i\cdot})$ is
\begin{equation}
    P\left(\mathbf{N}_t(A_{i\cdot}) \mid \{\pi_t(A_{ij})\}, N_{it} \right) = \frac{1}{C_0(\boldsymbol{\omega}_t(\mathbf{A}_{i\cdot}))} \prod_{j=1}^{n_i^{(c)}} \binom{P_t(A_{ij})}{N_t(A_{ij})}  \left( \omega_t(A_{ij}) \right)^{N_t(A_{ij})},
    \label{eq:fisher}
\end{equation}
where $\sum_{j=1}^{n_i^{(c)}} N_t(A_{ij}) = N_{it}$, $\omega_t(A_{ij}) = \frac{\pi_t(A_{ij})}{1-\pi_t(A_{ij})}$ are the atom-level odds of misuse within county $i$, and  
$C_0(\boldsymbol{\omega}_t(\mathbf{A}_{i\cdot}))$ is the normalizing constant. 


We assume atom-level OMP relates to covariates and a spatio-temporal random effect,
\begin{equation}
    \text{logit}\left(\pi_t(A_{ij}) \right) = \log\left( \omega_t(A_{ij})\right) =  \mathbf{W}_t^{(A)}(A_{ij})' \boldsymbol{\gamma}^{(A)} + \zeta_t(A_{ij}),
    \label{eq:atompi}
\end{equation}
where $\mathbf{W}_t^{(A)}(A_{ij})$ is a vector of covariates, $\boldsymbol{\gamma}^{(A)}$ the corresponding regression coefficients, and $\zeta_t(A_{ij})$ a spatio-temporal random effect. As was done previously, to reduce the dimensionality, we incorporate spatial basis functions and assume 
    $\zeta_t(A_{ij}) = \mathbf{B}(A_{ij})' \boldsymbol{\theta}_t,$
where $\mathbf{B}(A_{ij})$ is a $q$-dimensional vector of Moran's I basis functions at atom $A_{ij}$. The vector $\boldsymbol{\theta}_t$ is a $q$-dimensional vector of coefficients for the basis functions, and the coefficients can change over time. We account for temporal dependence by assuming an autoregressive model of order one for each of the $i=1,...,q$ components,
\begin{align*}
    \theta_{1,i} &\sim N\left(0, \tau^2_{A,i}\right) \, \, \text{ for } t=1\\
    \theta_{t,i} &\sim N\left(\phi_{A,i} \theta_{t-1,i}, \tau^2_{A,i}\right) \, \, \text{ for } t \geq 1.
\end{align*}

We assign prior distributions for all remaining parameters. All regression coefficients, $\bfbeta^{(o)}$ and $\boldsymbol{\gamma}^{(A)}$ are independently assigned mean-zero normal distributions with standard deviations of 5. The temporal autocorrelation parameters, $\phi_i^{(o)}$ and $\phi_{A,i}$, and the random effect variance parameters, $\tau^2_{o,i}$ and $\tau^2_{A,i}$, are assigned uniform distributions over $(0,1)$. Note that we chose these informative priors for the variances of the random effects to overcome identifiability challenges between the random effect and intercept parameters.

\subsection{Two-Stage MCMC for Bayesian Inference}
\label{ss:comp}


The full Bayesian model combines all model components specified in sections \ref{ss:BIA} and \ref{ss:model} in a hierarchical manner. Inference for the MIA model requires sampling from a high dimensional posterior distribution involving latent counts at both the county and atom levels. In addition to challenges with spatial dimensionality, a key computational challenge arises from the constrained structure of the atom-level counts, which must be nonnegative and sum to the county-level totals. We propose a two-stage compositional MCMC strategy that exploits the factorization of the posterior distribution and novelly combines efficient sampling algorithms with parallel computing. 

Let $\boldsymbol{\theta}_C$ denote the vector of all parameters (excluding the latent counts of interest) in the county-level model of Section \ref{ss:BIA}. That is, $$\boldsymbol{\theta}_C = \left( \{\mu_t^{(k)}\}, \{ \boldsymbol{\beta}^{(k)} \}, \{\mathbf{f}_{it}^{(k)}\}, \{\epsilon_{it}^{(k)} \}, \{\boldsymbol{\Sigma}_{\epsilon} \}, \{\tau^2_k  \}, \{\phi_k \}, \beta_0^{\mu}, \beta_1^{\mu}, \boldsymbol{\gamma}, \{u_{it}\}, \phi_u, \tau^2_u, \{v_{it} \}, \sigma^2_v   \right).$$ Similarly, let $\mathbf{\theta}_Z$ denote the parameters from the ZCTA-level indicator model  $$\boldsymbol{\theta}_Z = \left( \{\boldsymbol{\beta}^{(o)}\}, \{f_t^{(o)}(B_{\ell j}) \}, \{ \boldsymbol{\omega}^{(o)}_t \}, \{\phi_i^{(o)} \}, \{ \tau^2_{o,i} \}\} \right),$$ $\boldsymbol{\theta}_A = \left(\boldsymbol{\gamma}^{(A)}, \{\boldsymbol{\theta}_t\} \right)$ the vector of parameters from the atom-level regression model, and $\boldsymbol{\theta}_H = \left(\{\boldsymbol{\phi}_{A,i} \}, \{\boldsymbol{\tau}^2_{A,i} \} \right)$ the remaining hyperparameters from the atom-level model. 
Below, we will use the bracket notation such that $\left[A \mid B \right]$ denotes the probability function of $A$ conditional on $B$. The posterior distribution is then of the form 
\begin{equation}
\begin{aligned}
    \left[ \{N_{it}\},\right.&\left.\{N_t(A_{ij})\}, \boldsymbol{\theta}_C, \boldsymbol{\theta}_A, \boldsymbol{\theta}_Z, \boldsymbol{\theta}_H  \mid  \{Y_{it}^{(k)}\}, \{Z_{\ell t}^{(o)}\}, \mathbf{S}  \right] \\ 
    &\propto \left[ \{Y_{it}^{(k)}\},  \mathbf{S} \mid \{N_{it}\},  \boldsymbol{\theta}_C \right] \left[ \{Z_{\ell t}^{(o)}\} \mid  \{N_t(A_{ij})\}, \boldsymbol{\theta}_Z  \right] \\ &\hspace{1cm} \times \left[ \{N_t(A_{ij})\} \mid  \{N_{it}\}, \boldsymbol{\theta}_A  \right] \left[ \boldsymbol{\theta}_A \mid \boldsymbol{\theta}_H \right] \left[\boldsymbol{\theta}_Z\right] \left[\boldsymbol{\theta}_H \right] \left[  \{N_{it}\} \mid \boldsymbol{\theta}_C \right] \left[  \boldsymbol{\theta}_C \right] \\
    & \propto \left[ \{N_t(A_{ij})\}, \boldsymbol{\theta}_A,\boldsymbol{\theta}_Z,\boldsymbol{\theta}_H \mid \{N_{it}\}, \{Z_{\ell t}^{(o)}\}  \right] \times \left[ \{N_{it}\}, \boldsymbol{\theta}_C \mid  \{Y_{it}^{(k)}\} , \mathbf{S} \right] .
    \end{aligned}
    \label{eq:posterior}
\end{equation}
This implies the full posterior distribution is the product of the county-level posterior, $\left[ \{N_{it}\}, \boldsymbol{\theta}_C \mid  \{Y_{it}^{(k)}\} , \mathbf{S} \right]$ and the atom-level distribution $\left[ \{N_t(A_{ij})\}, \boldsymbol{\theta}_A,\boldsymbol{\theta}_Z,\boldsymbol{\theta}_H \mid \{N_{it}\}, \{Z_{\ell t}^{(o)}\}  \right].$ 

The constraint $N_{it} = \sum_{j=1}^{n_i^{(c)}} N_t(A_{ij})$ makes efficient simulation of \eqref{eq:posterior} difficult with standard MCMC algorithms. However, the structure of the posterior distribution lends itself to a two-stage compositional MCMC approach in which we simulate from the full posterior through the following two stages:
\begin{enumerate}
    \item Simulate $M$ draws, $\left(\{N_{it}^{(m)}\}, \boldsymbol{\theta}_C^{(m)}  \right)$ for $m=1,...,M$, from the county-level posterior.  
    \item For each draw $\left(\{N_{it}^{(m)}\}, \boldsymbol{\theta}_C^{(m)}  \right)$ from the county-level posterior, run a Metropolis-within-Gibbs MCMC algorithm to simulate from $ \left[ \{N_t(A_{ij})\}, \boldsymbol{\theta}_A,\boldsymbol{\theta}_Z,\boldsymbol{\theta}_H \mid \{N_{it}^{(m)}\}, \{Z_{\ell t}^{(o)}\}  \right]$. Save the last draw from the MCMC algorithm as $\left( \{N_t(A_{ij})^{(m)}\}, \boldsymbol{\theta}_A^{(m)},\boldsymbol{\theta}_Z^{(m)},\boldsymbol{\theta}_H^{(m)}  \right)$.
\end{enumerate}
To generate the samples from the county-level model in stage 1, we follow the approach of \cite{hepler_integrated_2023} and implement an adaptive Metropolis-within-Gibbs Sampler implemented using the R package NIMBLE \citep{nimble}. We use an automated factor slice sampler to block sample the $T$-dimensional vector of latent counts across all years for each county. Once the draws from the county-level model are obtained, we perform stage 2 in parallel for each $m=1,...,M$ to simulate from 
\begin{align}
    \left[ \{N_t(A_{ij})\}, \boldsymbol{\theta}_A,\boldsymbol{\theta}_Z,\boldsymbol{\theta}_H \right.&\left. \mid \{N_{it}^{(m)}\}, \{Z_{\ell t}^{(o)}\}  \right] \propto  \left[ \{Z_{\ell t}^{(o)}\} \mid  \{N_t(A_{ij})\}, \boldsymbol{\theta}_Z  \right] \nonumber\\  
    & \times \left[ \{N_t(A_{ij})\} \mid  \{N_{it}^{(m)}\}, \boldsymbol{\theta}_A  \right] \left[\boldsymbol{\theta}_A \mid \boldsymbol{\theta}_H  \right] \left[ \boldsymbol{\theta}_Z \right] \left[\boldsymbol{\theta}_H   \right].
    \label{eq:Natomfull}
\end{align}

In stage 2, Metropolis-within-Gibbs steps are used for each of the remaining parameters; we describe the update for each in what follows. First, note that a Metropolis-Hastings (MH) update of $\boldsymbol{\theta}_A$ is complicated due to the intractability of the normalizing constant in \eqref{eq:fisher}.  Observe, the full conditional distribution of $\boldsymbol{\theta}_A=\left[ \boldsymbol{\gamma}, \{\boldsymbol{\theta}_t\} \right]$ is
\begin{align*}
    \left[ \boldsymbol{\theta}_A \mid \cdot \right] &\propto  \left[ \{N_t(A_{ij})\} \mid  \{N_{it}^{(m)}\}, \boldsymbol{\theta}_A  \right] \left[\boldsymbol{\theta}_A \mid \boldsymbol{\theta}_H  \right] \\
    &= \prod_{i,t} \left\{ \frac{1}{C_0(\boldsymbol{\omega}_t(\mathbf{A}_{i\cdot}))} \prod_{j=1}^{n_i^{(c)}} \binom{P_t(A_{ij})}{N_t(A_{ij})}  \left( \omega_t(A_{ij}) \right)^{N_t(A_{ij})} \right\} \times  \left[\boldsymbol{\theta}_A \mid \boldsymbol{\theta}_H  \right].
    \label{eq:test}
\end{align*}
Suppose we simulate a proposed value $\boldsymbol{\theta}_A^* \sim q(\cdot \mid \boldsymbol{\theta}_A)$, and let $\omega_t(A_{ij})^*$ denote the corresponding odds. 
The MH acceptance probability is then $\min\{1, R\}$ where
\begin{equation}
    \begin{aligned}
        R &= \frac{\prod_{i,t} \left\{ \frac{1}{C_0(\boldsymbol{\omega}_t(\mathbf{A}_{i\cdot})^*)} \prod_{j=1}^{n_i^{(c)}} \binom{P_t(A_{ij})}{N_t(A_{ij})}  \left( \omega_t(A_{ij})^* \right)^{N_t(A_{ij})} \right\} \times  \left[\boldsymbol{\theta}_A^* \mid \boldsymbol{\theta}_H  \right]}{\prod_{i,t} \left\{ \frac{1}{C_0(\boldsymbol{\omega}_t(\mathbf{A}_{i\cdot}))} \prod_{j=1}^{n_i^{(c)}} \binom{P_t(A_{ij})}{N_t(A_{ij})}  \left( \omega_t(A_{ij}) \right)^{N_t(A_{ij})} \right\} \times  \left[\boldsymbol{\theta}_A \mid \boldsymbol{\theta}_H  \right]} \times \frac{q(\boldsymbol{\theta}_A \mid \boldsymbol{\theta}_A^*)}{q(\boldsymbol{\theta}_A^*\mid \boldsymbol{\theta}_A)} 
        \label{eq:MHR}
    \end{aligned}
\end{equation}
The ratio of FNCH normalizing constants, $ \prod_{i,t} \frac{C_0(\boldsymbol{\omega}_t(\mathbf{A}_{i\cdot}))}{C_0(\boldsymbol{\omega}_t(\mathbf{A}_{i\cdot})^*)},$ is computationally prohibitive to compute as it requires summation over all possible vectors of atom-level counts. The \texttt{BiasedUrn} R package has built-in functionality to compute the probability function, but it is extremely computationally expensive when $N_{it}$ is large, which makes it prohibitive for use within an MCMC algorithm  \citep{ fog2008calculation, BiasedUrn}. 

Recall the FNCH distribution is the conditional distribution that would be obtained if each $N_t(A_{ij})$ was independently a Binomial random variable with population $P_t(A_{it})$ and odds of success $\omega_t(A_{ij})$, but conditioned on the constraint that $\sum_{j=1}^{n_i^{(c)}} N_t(A_{ij}) = N_{it}$. When $N_{it}$ is small relative to the county population, as it is in our application, the contribution of the normalizing constant in \eqref{eq:fisher} is negligible and likelihood ratios are well approximated by those obtained under independent Binomial models \citep{harkness1965properties}. Thus, $R \approx \hat{R}$, and we approximate the MH update by accepting $\boldsymbol{\theta}_A^*$ with probability $\min\{1, \hat{R} \}$, where
\begin{equation}
    \begin{aligned}
        \hat{R} &= \frac{\prod_{i,t} \left\{ \prod_j  \binom{P_t(A_{ij})}{N_t(A_{ij})} \left( \omega_t(A_{ij}^*) \right)^{N_t(A_{ij})} \left(1+\omega_t(A_{ij}^*) \right)^{-P_t(A_{ij})} \right\} \times  \left[\boldsymbol{\theta}_A^* \mid \boldsymbol{\theta}_H  \right]}{\prod_{i,t} \left\{ \prod_j  \binom{P_t(A_{ij})}{N_t(A_{ij})} \left( \omega_t(A_{ij}) \right)^{N_t(A_{ij})} \left(1+\omega_t(A_{ij}) \right)^{-P_t(A_{ij})} \right\} \times  \left[\boldsymbol{\theta}_A \mid \boldsymbol{\theta}_H  \right]} \times \frac{q(\boldsymbol{\theta}_A \mid \boldsymbol{\theta}_A^*)}{q(\boldsymbol{\theta}_A^*\mid \boldsymbol{\theta}_A)}.
        \label{eq:MHRhat}
    \end{aligned}
\end{equation}

We propose a block random walk MH update for the $n_i^{(c)}$-dimensional vector $\mathbf{N}_t(\mathbf{A}_{i\cdot})$ for each $i, t$ that enforces the constraint $\sum_{j=1}^{n_i^{(c)}} N_t(A_{ij}) = N_{it}^{(m)}$. The idea is to randomly select two atoms within county $i$, a source and a destination, and move a subset of counts from the source to the destination. Since both atoms are nested within the same county, the county-level sum is preserved. This sampler was motivated by and adapted from the multinomial random walk sampler in NIMBLE and is described in more detail below.

Let $\mathbf{N}_t(\mathbf{A}_{i\cdot})$ denote the current state at a given iteration of the MCMC algorithm. A candidate $\mathbf{N}_t(\mathbf{A}_{i\cdot})^*$ is generated as follows.
\begin{enumerate}
    \item Select a source index $j$ uniformly from the set of atoms within county $i$ with positive counts, given by $\mathcal{S}(\mathbf{N}_t(A_{i\cdot}))$. That is, if $\mathcal{S}(\mathbf{N}_t(A_{i\cdot})) = \{j : N_t(A_{ij})>0\}$, then we select atom $A_{ij}$ as the source with probability      $P(j |\mathbf{N}_t(A_{i\cdot})) = \frac{1}{\left|\mathcal{S}(\mathbf{N}_t(A_{i\cdot}))\right|}.$
    \item Select a destination index $k$ uniformly from $\{1,...,n_i^{(c)} \}$ such that $P(k |\mathbf{N}_t(A_{i\cdot})) = \frac{1}{n_i^{(c)}}$
    \item Randomly select the move size $\tilde{n}$ uniformly from $\{1,2,...,\min(\tilde{M}, N_t(A_{ij})) \}$ where $\tilde{M}$ is a tuning parameter. We chose $\tilde{M} = \lfloor N_{it}/20 \rfloor$ in our implementation.
    \item Propose $N_t(A_{ij})^* = N_t(A_{ij})-\tilde{n}$ and $N_t(A_{ik})^* = N_t(A_{ik})+\tilde{n}$, and all other atoms within the county remain unchanged.
\end{enumerate}
The proposal distribution is then
\begin{equation}
    q\left(\mathbf{N}_t(A_{i\cdot})^* | \mathbf{N}_t(A_{i\cdot})  \right) =  \frac{1}{\left|\mathcal{S}(\mathbf{N}_t(A_{i\cdot}))\right|} \cdot \frac{1}{n_i^{(c)}} \cdot \frac{1}{\min(\tilde{M}, N_t(A_{ij}))},
    \label{eq:proposal}
\end{equation}
and the proposed vector is accepted with probability
\begin{equation*}
    \alpha(\mathbf{N}_t(A_{i\cdot}), \mathbf{N}_t(A_{i\cdot})^*) = \min\left(1, \frac{\left[\mathbf{Z}_t \mid \mathbf{N}_t(\mathbf{B})^*, \boldsymbol{\theta}_A \right] }{\left[\mathbf{Z}_t \mid \mathbf{N}_t(\mathbf{B}), \boldsymbol{\theta}_A \right] } \cdot \frac{\left[\mathbf{N}_t(\mathbf{A}_{i\cdot})^* \mid N_{it}, \boldsymbol{\theta}_A \right]}{\left[\mathbf{N}_t(\mathbf{A}_{i\cdot}) \mid N_{it}, \boldsymbol{\theta}_A \right]} \cdot \frac{q\left(\mathbf{N}_t(\mathbf{A}_{i\cdot}) | \mathbf{N}_t(\mathbf{A}_{i\cdot})^*  \right)}{q\left(\mathbf{N}_t(\mathbf{A}_{i\cdot})^* | \mathbf{N}_t(\mathbf{A}_{i\cdot})  \right)}   \right).
\end{equation*}
Note that based on this proposal distribution, only two elements of  $\mathbf{N}_t(\mathbf{A}_{i\cdot})$ and $\mathbf{N}_t(\mathbf{A}_{i\cdot}^*)$ differ, which corresponds to atoms $A_{ij}$ and $A_{ik}$, and they differ such that  $N_t(A_{ij})^* = N_t(A_{ij})-\tilde{n}$ and $N_t(A_{ik})^* = N_t(A_{ik})+\tilde{n}$. Thus, the ratio of FNCH distributions can easily be computed as it is equivalent to what would have been obtained had the atom-level counts been modeled with independent Binomial distributions. That is,
\begin{align*}
     \frac{\left[\mathbf{N}_t(\mathbf{A}_{i\cdot})^* \mid N_{it}, \boldsymbol{\theta}_A \right]}{\left[\mathbf{N}_t(\mathbf{A}_{i\cdot}) \mid N_{it}, \boldsymbol{\theta}_A \right]} &= \left\{\prod_{r=0}^{\tilde{n}-1} \frac{N_t(A_{ij})-r}{P_t(A_{ij})-N_t(A_{ij})+r+1} \cdot \frac{P_t(A_{ik})-N_t(A_{ik})-r}{N_t(A_{ik})+r+1}  \right\} \\ &\times \left( \frac{\pi_t(A_{ik})/(1-\pi_t(A_{ik}))}{\pi_t(A_{ij})/(1-\pi_t(A_{ij}))} \right)^{\tilde{n}}.
\end{align*}

Without loss of generality, let $\ell_1$ and $\ell_2$ denote the ZCTAs where the source and destination atom reside, respectively, indexed such that $B_{\ell_1, j} = A_{i,j}$ and $B_{\ell_2, k} = A_{i,k}$. Let $\lambda_t^{(o)}(\mathbf{B}_{\ell,\cdot}) = \sum_i N_t(B_{\ell,i}) q_t^{(o)}(B_{\ell, i})$ be the mean of the ZCTA-level model in \eqref{eq:naloxonemodel} for ZCTA $\ell$ at time $t$, and $\lambda_t^{(o)}(\mathbf{B}_{\ell,\cdot})^* = \sum_i N_t(B_{\ell,i})^* q_t^{(o)}(B_{\ell, i})$ the corresponding mean given the proposed vector of latent counts ($N_t(B_{\ell,i})^*=N_t(B_{\ell,i})$ for all but the source and destination atoms). Then the ratio of the ZCTA-likelihood simplifies to
\begin{align*}
    \frac{\left[\mathbf{Z}_t \mid \mathbf{N}_t(\mathbf{B}^*), \boldsymbol{\theta}_A \right] }{\left[\mathbf{Z}_t \mid \mathbf{N}_t(\mathbf{B}), \boldsymbol{\theta}_A \right] } &= \prod_{o} \exp \left(\tilde{n}(q_t^{(o)}(B_{\ell_1,j}) -q_t^{(o)}(B_{\ell_2,k}) )  \right)\\ &\times \left( 1 - \frac{\tilde{n}q_t^{(o)}(B_{\ell_1,j})}{\lambda_t^{(o)}(\mathbf{B}_{\ell_1, \cdot})} \right)^{Z_{\ell_1}}  \left( 1 + \frac{\tilde{n}q_t^{(o)}(B_{\ell_2,k})}{\lambda_t^{(o)}(\mathbf{B}_{\ell_2, \cdot})} \right)^{Z_{\ell_2}}
\end{align*}
since $\lambda_t^{(o)}(\mathbf{B}_{\ell_1,\cdot})^* = \lambda_t^{(o)}(\mathbf{B}_{\ell_1,\cdot}) - \tilde{n} q_t^{(o)}(B_{\ell_1, j})$ and $\lambda_t^{(o)}(\mathbf{B}_{\ell_2,\cdot})^* = \lambda_t^{(o)}(\mathbf{B}_{\ell_2,\cdot}) + \tilde{n} q_t^{(o)}(B_{\ell_1, j})$. Thus, the acceptance probability, $\alpha\left( \mathbf{N}_t(\mathbf{A}_{i \cdot}), \mathbf{N}_t(\mathbf{A}_{i \cdot}^*) \right)$, can easily be computed exactly.

The remaining parameters, $\boldsymbol{\theta}_Z$ and $\boldsymbol{\theta}_H$, can be updated using standard random walk Metropolis-Hastings updates. We implement this second stage Metropolis-within-Gibbs algorithm in NIMBLE in parallel for each of the $m=1,...,M$ draws from the stage one output. To code the atom-level models, we take advantage of the fact that the MH acceptance probability for the latent atom-level counts and the \textit{approximate} MH acceptance probability for the atom-level model parameters, $\boldsymbol{\theta}_A$ are equivalent to what would be obtained had we assumed Binomial distributions instead of the FNCH. Default samplers are used for all parameters in $\boldsymbol{\theta}_A$, $\boldsymbol{\theta}_Z$, and $\boldsymbol{\theta}_H$, but we customize the sampler for the latent atom-level counts to implement the block random walk MH algorithm proposed above. For each $m=1,...,M$, we save the last iteration's values and denote these by $\boldsymbol{\theta}_A^{(m)}$, $\boldsymbol{\theta}_Z^{(m)}$, $\boldsymbol{\theta}_H^{(m)}$, and $\mathbf{N}_t(\mathbf{A}_{i\cdot})^{(m)}$. Concatenating these with the stage one output yields samples, $\left(\{N_{it}^{(m)}\}, \{N_t(A_{ij})^{(m)}\}, \boldsymbol{\theta}_C^{(m)}, \boldsymbol{\theta}_A^{(m)} \right)$ for $m=1,...,M$, approximately from the posterior. All code to implement the stage one and stage two MCMC algorithms is available in the SM.

\section{Simulation Study}
\label{sec:sim}

We performed a simulation study to investigate how well the MIA model recovers atom-level latent OMP. Since the county-level model has previously been shown to accurately recover county-level counts, $N_{it}$ \citep{hepler_integrated_2023}, this simulation will focus on the multiscale model of Section \ref{ss:model} and implement the second stage of the MCMC algorithm, assuming a fixed set of county-level counts. 

Our simulation mimics the real-data application by using the same spatial structure and covariates. We assume the county-level counts $N_{it}$ are known, and we fix those to be equal to the posterior median of the real-data analysis from Section \ref{sec:app}, shown in SM Figure 19. We assume atom-level OMP depends on two covariates, namely the logit of the proportion of the population that is unemployed and the logit of the proportion of the population that has at least a Bachelor's degree. The assumed values of the regression coefficients are $\boldsymbol{\gamma}^{(A)} = \left(-1.5, 0.75, -0.5 \right)'$, and the coefficients of the basis functions in the spatio-temporal random effect were generated from an AR(1) model assuming temporal autocorrelation of 0.95 and standard deviation of 0.75. We simulated $H=100$ sets of atom-level counts using the \texttt{BiasedUrn} R package to simulate from the FNCH distribution \citep{BiasedUrn}.

For each simulated set of atom-level counts, we generated a single ($o=1$) ZCTA-level indicator according to the model in Section \ref{ss:zctamod}. We assume the rate, $q_t^{(o)}(B_{\ell j})$, depends on indicators of metropolitan and micropolitan. The regression coefficients are $\boldsymbol{\beta}^{(o)} = (-3.0, 1.0, 0.4)'$. The spatio-temporal random effect is simulated assuming coefficients of the basis functions follow an AR(1) model with temporal autocorrelation of 0.9 and standard deviation of 1.75. Maps of two of the simulated data sets can be found in SM Figures 8-13.

We fit the ZCTA and atom-level portions of the MIA model from Section \ref{ss:model} for each of the $H=100$ simulated ``naloxone'' ZCTA data sets. We implement the second stage of the MCMC algorithm described in Section \ref{ss:comp} to simulate from the conditional posterior $\left[ \{N_t(A_{ij})\}, \boldsymbol{\theta}_A,\boldsymbol{\theta}_Z,\boldsymbol{\theta}_H \mid \{N_{it}\}, \{Z_{\ell t}^{(o)}\}  \right]$. We also compare our model performance to two naive approaches to downscaling the county-level counts. The first comparison generates atom-level estimated counts by assuming the county-level counts are downscaled proportionally to the atom-level population. That is, the naive population-based estimates are computed as $\hat{N}_t^{(P)}(A_{ij}) = N_{it} \times \frac{P_t(A_{ij})}{P_{it}}$. 
The second comparison generates atom-level counts by downscaling the county-level counts in proportion to the count of naloxone administrations of the atoms within that county. Note that since naloxone counts are ZCTA-level, we first downscale to the atom level in proportion to population. That is, the naloxone-based naive estimates are $\hat{N}^{(Z)}_t(A_{ij}) = N_{it} \times \frac{\hat{Z}_t(A_{ij})}{\sum_j \hat{Z}_t(A_{ij}) },$ where $\hat{Z}_t(A_{ij}) = Z_{\ell,t} \times \frac{P_t(B_{\ell k})}{\sum_k P_t(B_{\ell k})}$ assuming ZCTA $\ell$ is the ZCTA that contains atom $A_{ij}$.



\subsection{Simulation Results}

The MCMC algorithms were implemented in parallel for each of the 100 simulated data sets. Each MCMC was run for 100,000 iterations using a burn-in of 50,000 iterations and thinning the output by storing every 10th sample. The average computing time across the 100 MCMC algorithms was 6.91 hours. We assessed convergence by randomly selecting several of the simulations and checking trace plots of model parameters. Maps of the estimated atom-level OMP are shown in SM Figures 14-16 for two of the simulated data sets. These maps show that our model recovers the spatio-temporal trends in latent atom-level OMP, particularly so in the more populous central Ohio region.



\begin{table}[ht]
\centering
\begin{tabular}{lrrrrrrr}
  \hline
 & Model & Min. & 1st Qu. & Median & Mean & 3rd Qu. & Max. \\ 
  \hline
\multirow{3}{*}{\rotatebox[origin=c]{90}{Bias}} & Proposed & -0.30919 & -0.01030 & -0.00162 & -0.00281 & 0.00589 & 0.21318 \\ 
 & Naive Population & -0.50980 & -0.01269 & 0.00349 & -0.00323 & 0.01602 & 0.10157 \\ 
 & Naive Naloxone & -0.29507 & -0.01501 & -0.00263 & 0.00701 & 0.01314 & 1.42720 \\ 
   \hline
\multirow{3}{*}{\rotatebox[origin=c]{90}{RMSE}} & Proposed & 0.00040 & 0.00587 & 0.01161 & 0.01805 & 0.02248 & 0.34801 \\ 
&  Naive Population & 0.00049 & 0.01013 & 0.01922 & 0.02641 & 0.03154 & 0.51156 \\ 
 & Naive Naloxone & 0.00063 & 0.01175 & 0.02152 & 0.03679 & 0.03920 & 1.44106 \\ 
   \hline
\multirow{3}{*}{\rotatebox[origin=c]{90}{MAE}}& Proposed & 0.00032 & 0.00510 & 0.01015 & 0.01548 & 0.01923 & 0.30919 \\ 
&  Naive Population & 0.00039 & 0.00913 & 0.01777 & 0.02466 & 0.02962 & 0.50980 \\ 
 & Naive Naloxone & 0.00052 & 0.01010 & 0.01861 & 0.03268 & 0.03391 & 1.42720 \\ 
\end{tabular}
\caption{Summary statistics of the empirical bias, RMSE, and MAE for the proposed model and naive estimates. Note these statistics summarize over all atoms and years. }
\label{tab:SimBiasRMSE}
\end{table}

We compare estimated OMP under the proposed MIA model and the two comparison models using the empirical bias, root mean-squared error (RMSE), and median absolute error (MAE). Definitions of the metrics can be found in SM Section 4, and boxplots are shown in SM Figure 17. Table \ref{tab:SimBiasRMSE} provides summary statistics of the metrics for the three models. The first and third quartile, median, and mean empirical bias are all closer to zero for the MIA. The maximum empirical bias for the proposed model is larger than that of the naive population-based estimates, but substantially smaller than for the naive naloxone-based estimates, which can actually exceed one since total naloxone administrations can surpass the population size. The RMSE is systematically smaller for the MIA than for the naive estimates. Notably, 79.9\% of OMP estimates from the proposed model were within 0.01 of the true OMP, compared to just 60.9\% and 69.9\% of the estimates from the naive population- and naloxone-based downscaled estimates, respectively. While on average, the empirical bias and RMSE for the proposed model are small, we acknowledge that maximum values of 0.213 and 0.348 for these quantities are large for a prevalence estimate. SM Figure 18 shows these plotted against atom-level population. Unsurprisingly, the largest errors are in atoms with very small populations, where small changes in the estimated count $\hat{N}_t(A_{ij})$ have a large impact on OMP, $\hat{N}_t(A_{ij})/P_t(A_{ij})$. 



\section{Application}
\label{sec:app}


We fit the proposed MIA model to the Ohio data described in Section \ref{sec:data}. Briefly, we jointly model state-level survey estimates on OMP, county-level counts of opioid overdose deaths and treatment admissions, and ZCTA-level counts of EMS-administered naloxone (post-processed from ZIP-code level data as previously described). We fit the model using the two-stage compositional MCMC approach described in Section \ref{ss:comp}.

For the coarser county-level component of the model, we include covariates, $\mathbf{X}_{it}^{(k)}$, to model the county-level rates of death and treatment admissions among PWMO, given by $p_{it}^{(1)}$ and $p_{it}^{(2)}$ in equation \eqref{eq:ctydata}. For the death rate, we include an indicator of whether a major interstate intersects the county as well as indicators of the county being a high impact drug trafficking area and a metropolitan statistical area. For the treatment rate, we include indicators of the county being a health professional shortage area and a medically underserved area. To model the county-level OMP ratio, $\lambda_{it}$, we include standardized covariates on the proportion of the county in poverty, unemployed, with at least a high school degree, and on food stamps, as well as the rate of prescription opioids dispensed. As in \cite{hepler_integrated_2023}, we fit the county-level model using NIMBLE in R using  AF slice samplers to jointly update the $T$-dimensional vector of county-level counts, $\mathbf{N}_{i\cdot}$, for each county. We ran the MCMC algorithm for 1,000,000 iterations using a burn-in of 500,000 and storing every 50th sample, resulting in $M=10,000$ samples from the county-level posterior distribution. The MCMC algorithm took 9.48 hours to run on a single core of the DEAC HPC computing cluster \citep{WakeHPC}. 

The model for latent atom-level rate of naloxone among PWMO, $q_t^{(o)}(B_{\ell j})$, depends on atom-level covariates $\mathbf{X}_t^{(o)}(B_{\ell j})$, as shown in \eqref{eq:zctarate}. We include two indicators for whether the atom belongs to a ZCTA that is classified as `metropolitan' (RUCA code between 1-3) or `micropolitan' (RUCA code between 4-6). The primary quantities of interest are the atom-level OMP, $\pi_t(A_{iJ})$, modeled as in \eqref{eq:atompi}. For the application, we assume this depends on the logit-transformed covariates: proportion of the population that is unemployed and proportion with at least a bachelor's degree. Note that these covariates are only available at the ZCTA-level, and we assume the proportions are the same for all atoms within the ZCTA. The spatial random effects for the ZCTA-level naloxone model and the atom-level OMP model use $q=50$ Moran's I basis functions. 

We perform the second stage of the two-stage compositional MCMC approach in parallel for each of the $M=10,000$ samples from the first county-level stage.  We ran the algorithm for 100,000 total iterations. Note that we only need to store the last sample, but to assess convergence the second stage MCMC, we randomly selecting several of the $M$ runs to store all samples and inspect trace plots of model parameters. The average run time for the $M$ second-stage MCMC algorithms was 6.4 hours, and these were implemented in parallel.

\subsection{Results}

\begin{figure}[h]
    \centering
    \includegraphics[trim=2in .5cm 1.8in 1cm, clip, width=0.8\linewidth]{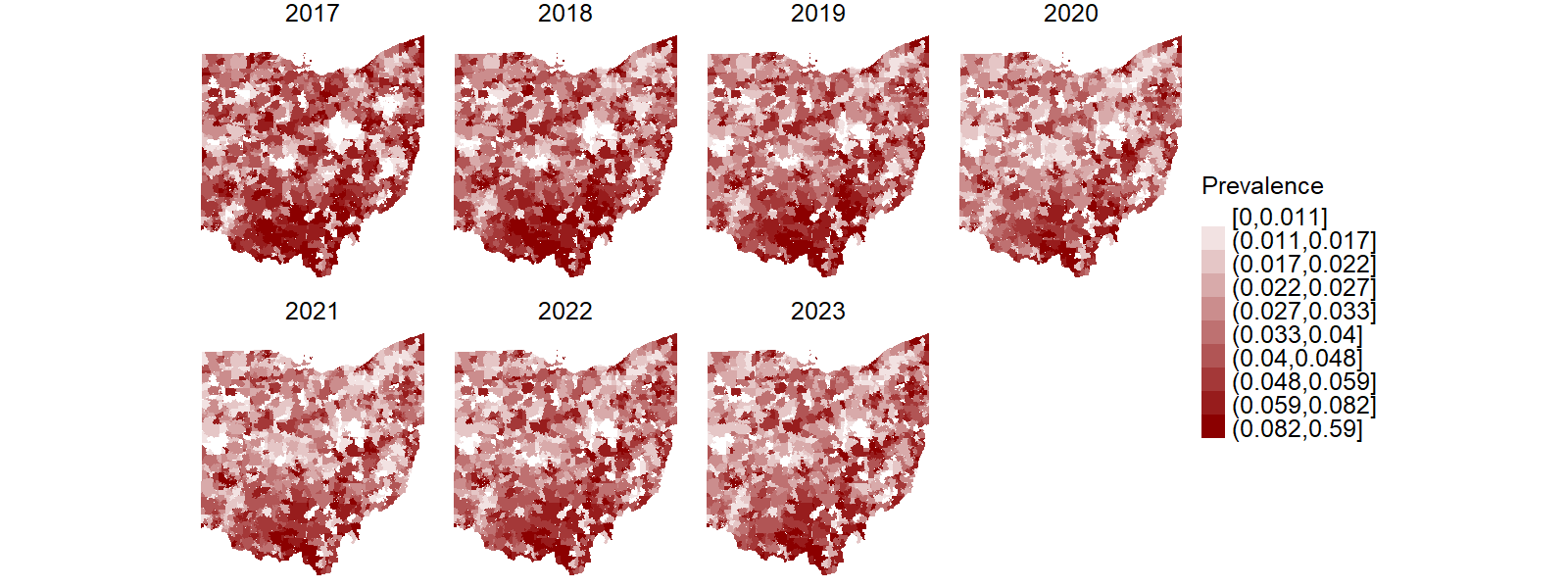}
    \caption{Estimated atom-level OMP binned into increments based on deciles.}
    \label{fig:AppAtomNbin}
\end{figure}

\begin{figure}[h]
    \centering
    \begin{subfigure}{.95\linewidth}
\includegraphics[trim=0in 1.35in 0in 1.2in, clip, width=0.9\linewidth]{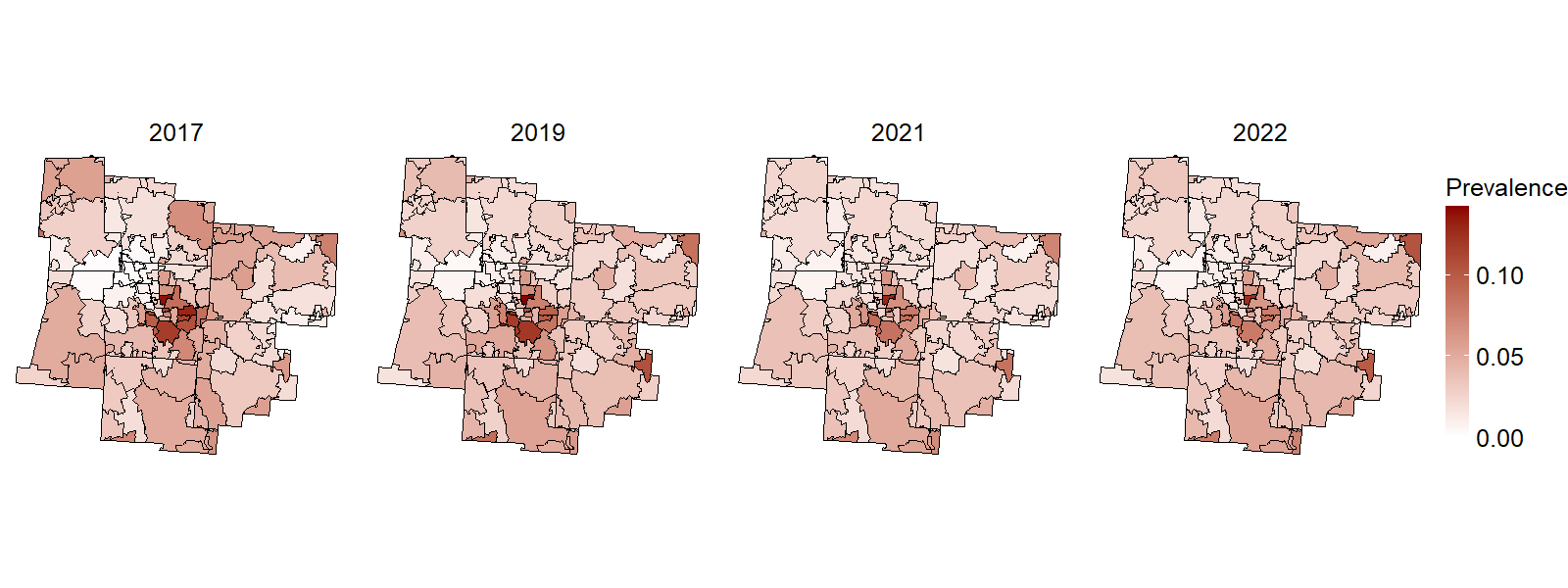} 
\label{fig:AppNhatCentral}
\caption{}        
    \end{subfigure}
        \begin{subfigure}{.98\linewidth}
        \includegraphics[trim=0in 1.4in 0in 1.2in, clip, width=0.9\linewidth]{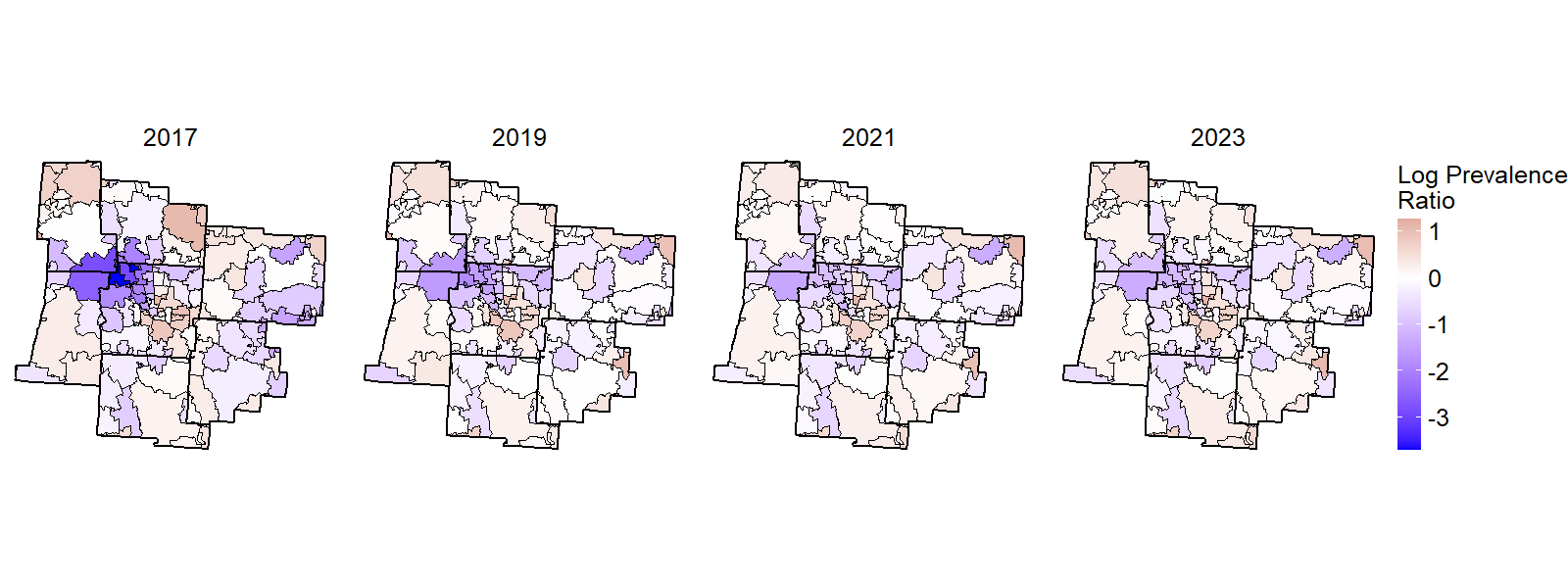} 
\label{fig:RelRateCentral} 
\caption{}        
    \end{subfigure}
    \caption{Estimated atom-level OMP, $N_t(A_{ij})/P_t(A_{ij})$, (Figure a) and log prevalence ratio of atom-level OMP relative to county-level OMP, $\log\left(\frac{N_t(A_{ij})/P_t(A_{ij})}{N_{it}/P_{it}} \right)$, (Figure b) zoomed in to atoms within central Ohio counties.}
    \label{fig:Central}
\end{figure}

\begin{figure}[h]
    \centering
    \begin{subfigure}{.95\linewidth}
\includegraphics[trim=0in 1.8in 0in 1.7in, clip, width=0.93\linewidth]{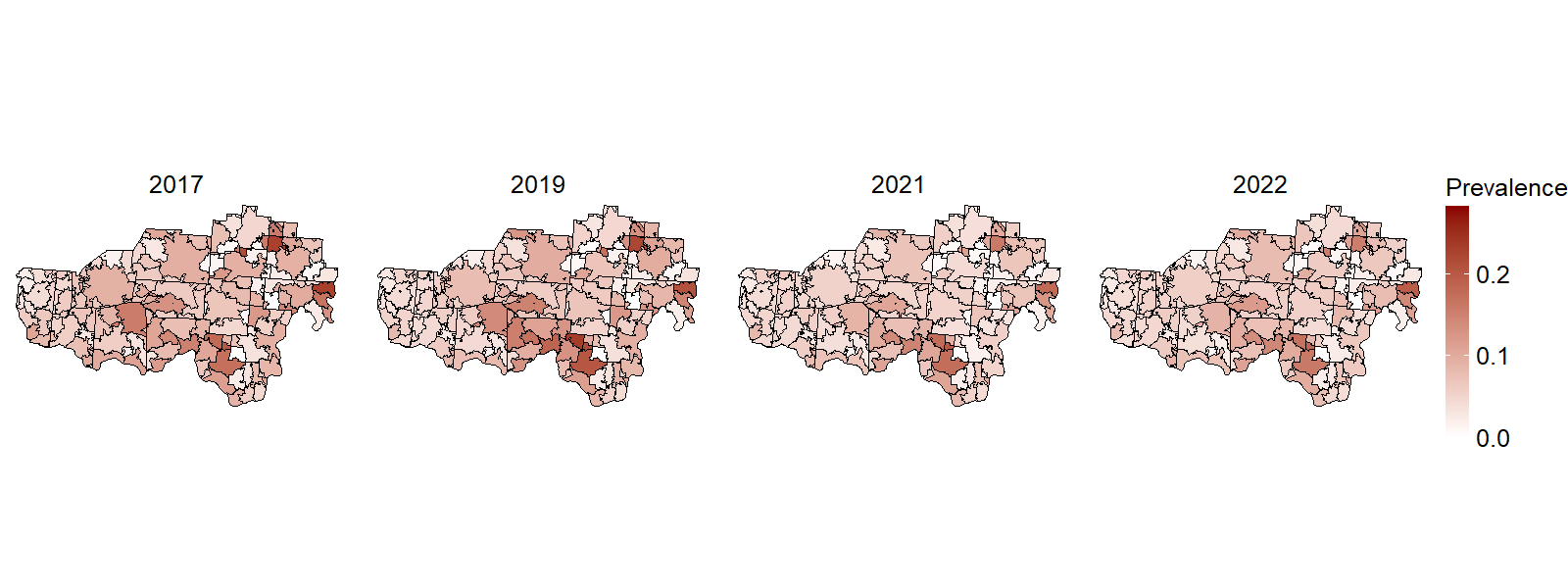} 
\label{fig:AppNhatSouthern}
\caption{}        
    \end{subfigure}
        \begin{subfigure}{.95\linewidth}
\includegraphics[trim=0in 1.9in 0in 1.65in, clip, width=0.95\linewidth]{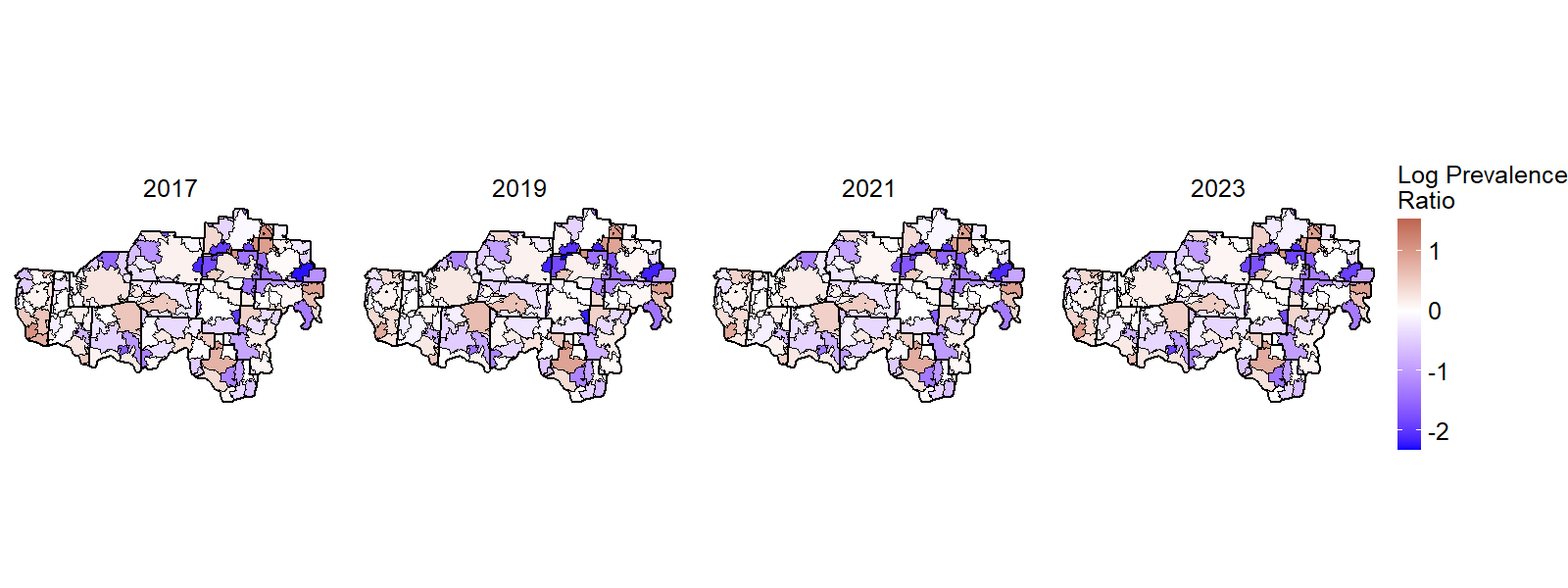} 
\label{fig:RelRateSouthern}
\caption{}        
    \end{subfigure}
    \caption{Estimated atom-level OMP (Figure a) and log prevalence ratio of atom-level OMP relative to county-level OMP (Figure b) zoomed in to atoms within southern Ohio.}
    \label{fig:Southern}
\end{figure}

A map of the estimated atom-level OMP, $\hat{N}_t(A_{ij})/P_t(A_{ij})$, is shown in Figure \ref{fig:AppAtomNbin}. For ease of visualization, the map shows OMP binned into ten categories created based on deciles. Estimated OMP with a continuous color scale is shown in SM Figure 25 with  standard deviations in SM Figure 26.  Figures \ref{fig:Central} and \ref{fig:Southern} show estimated OMP and the log  prevalence ratio relative to county OMP, zoomed to only include atoms contained within central and southern Ohio, respectively. Comparing these maps to the county-level map shown in SM Figure 19 yields interesting insight into the within-county heterogeneity that would be missed by a county-only analysis. For example, Franklin County in central Ohio had an estimated OMP of approximately 0.05 in 2017, decreasing to roughly 0.03 in 2023. However, the multiscale analysis indicates that some atoms within Franklin County have estimated OMP roughly three times higher, reaching values of approximately 0.15. At the county-level, Franklin County is in the middle tercile of opioid misuse prevalence, but there are atoms within Franklin county that are in the 99th percentile of OMP across the state. Scioto County in southern Ohio, considered one of the hardest hit counties in America for the opioid epidemic, has estimated prevalence as high as 0.12 in 2019. The prevalence ratios for atoms in this county range from approximately 0.1 to 1.8, indicating that while some atoms have much lower OMP than the county, other atoms are 80\% higher. The log of the prevalence ratio for the entire state is in SM Figure 27.



\begin{table}[ht]
\centering
\begin{tabular}{lrl}
  \hline
  Parameter & Mean & 95\% CI \\ 
  \hline
Intercept & 0.46 & (0.36, 0.61) \\ 
Unemployment & 2.65 & (2.45, 2.9) \\ 
College Degree & 0.75 & (0.71, 0.8) \\ 
   \hline
\end{tabular}
\caption{Posterior mean and 95\% credible intervals or prevalence ratios in the atom-level model of prevalence of people who misuse opioids. }
\label{tab:atomcoeff}
\end{table}

A key feature of the MIA is that regression coefficients in the atom-level model can be interpreted at the population-level, like any standard disease mapping model.  Table \ref{tab:atomcoeff} shows posterior means and 95\% credible intervals of the prevalence ratios in the atom-level model of OMP, $e^{\gamma^{(A)}}$. Note that the variables were logit-transformed, so this represents the multiplicative change in the odds for an increase of one on the logit scale. We estimate atom-level prevalence increases as the proportion of the population who is unemployed increases, and prevalence decreases at the proportion of the population with a college degree increases. We also estimated that atom-level rates of EMS naloxone administrations per PWMO were 77\% (95\% CI: 65\%, 88\%) higher in metropolitan compared to rural atoms (SM Table 3). Given we have more EMS naloxone administrations in more urban areas, we are able to estimate more heterogeneity in OMP, as naloxone administrations are observed in the smallest areal unit of the indicators. This also leads to estimates of OMP that shrink to the county OMP and are generally smoother in more rural areas where there are fewer naloxone administrations to inform heterogeneity in OMP across atoms. This is an appealing feature of the MIA as we are able to use finer scale information when it is available while shrinking to the county OMP when we lack sufficient information to estimate more heterogeneity. 

The MIA is also fully interpretable at the county level. A map of estimated county-level OMP is shown in SM Figure 19 with the posterior standard deviation in SM Figure 20. Estimated regression coefficients are in SM Table 2. We estimated positive associations of county-level OMP with the proportions in poverty, on food stamps, with at least a high school degree, and unemployment. We estimated a small negative association with the opioid prescribing rate, which is potentially partly attributable to the shift in the epidemic away from prescription opioids during the time period of this study. Maps of the estimated overdose death rate among PWMO and its standard deviation are shown in SM Figures 21-22. We estimated that the county-level odds of overdose death among PWMO were 21\% (95\% CI: 11\%, 31\%) higher in metropolitan areas compared to non-metropolitan areas. Maps of the estimated opioid treatment admissions among PWMO and its standard deviation are shown in SM Figures 23-24. We did not have strong evidence of associations of covariates with treatment admissions. Finally, annual estimates of state-level OMP, derived from the multi-year NSDUH estimates are shown in SM Figure 28 and exhibit a decreasing trend over time.

\section{Discussion}
\label{sec:discuss}

In this paper, we developed the multiscale integrated abundance model (MIA), which builds on the BIA of \citet{hepler_integrated_2023} to estimate OMP at the sub-county level while remaining simultaneously interpretable at the county level. The MIA is designed so that each layer of the model is epidemiologically interpretable as areal regression models for disease mapping. Latent sub-county OMP estimates are generated using a novel approach based on the FNCH which incorporates constraints requiring sub-county atom-level counts within a county to sum to the county total and to not exceed the atom total population. To make the model fitting computationally feasible, we developed a two-stage compositional MCMC algorithm with an approximate MH update to overcome the computational intractability of the normalizing constant associated with the FNCH. A simulation study verified that the MIA could recover atom-level prevalence estimates and outperformed naive competing approaches. We applied the MIA to data from Ohio to estimate county- and sub-county-level OMP to further characterize heterogeneity and provide estimates that can improve local targeting of public health services and interventions.

Estimates from the MIA make significant contributions to the epidemiological understanding and public health response to the opioid epidemic. Previously, statewide, population-based estimates were primarily limited to the county level. More local analysis were typically concentrated within specific cities or counties \citep{wesson_bayesian_2018,mcneely_estimating_2012}. With the MIA, we are now able to provide county and sub-county estimates of OMP jointly for all counties in the state. Because of constraints on sub-county estimates, we ensure that estimates within the model are internally consistent so one can draw the same conclusions regardless of which spatial unit is being considered.  This enables MIA estimates to be used by state officials to help guide resources to counties and to be used by counties to help optimize the response within their jurisdiction. In addition, the model enables the incorporation of covariates and retains standard epidemiological interpretations of parameters which facilitates exploration of other social environmental questions of interest and eases the integration of this work into practice. The general MIA framework can also be applied to other challenging to measure health outcomes to integrate multiple sources of data to generate small area estimates.

The MIA modeling framework opens up a range of potential future research. The availability of sub-county-level OMP estimates enable a more granular examination of questions related to access and gaps in the availability of health services. Prior to the MIA, questions were limited to the county-level which served as crude approximations to local accessibility of services. Future methodological research could continue to explore ways to estimate OMP on finer geographic scales, should such indicator data be available. Extensions could include connections to and adaptations of the work of \citet{bradley2016bayesian} which would enable estimation at any geographical unit of choice. This would be particularly advantageous as different health departments and service organizations may have different priorities and require different spatial units of analysis. However, a unified framework across units of analysis would be highly desirable so that resulting inferences and actions are consistent across agencies and not dependent on the choice of the unit of analysis.

Like any model for estimating a hidden population, potential departures from assumptions represent limitations. Estimates of OMP counts depend on the NSDUH survey reflecting a reasonable scale for OMP. However, prevalence ratios are robust to the quality of the NSDUH survey data. We also assume that all individuals counted in each indicator are members of the population of interest, PWMO, and like any surveillance data, a chance of misclassification (i.e., false positives) may exist. It is also unlikely that all EMS naloxone administrations occur in the home ZIP code of the patient, although we believe this still serves as a reasonable sub-county indicator. Such spatial misattribution is a topic for future research. Spatial resolution of the estimates are limited by the availability of finer scale indicators; however, the 2-stage computing approach enables the incorporation of new indicators as they become available. Finally, we had to use an approximate MH update for $\theta_A$ parameters for computational feasibility; however, the approximation is likely reasonable as OMP is relatively low.

In conclusion, the MIA integrates data from multiple sources at multiple levels of aggregation of spatially misaligned units to produce multiscale estimates of OMP while quantifying their associated uncertainty. The multiscale estimates enable decision-makers at the state and local levels of the public health system to have estimates that are relevant to their jurisdictions that are jointly estimated and consistent across levels of aggregation. Estimates from the MIA are a major advancement in understanding local epidemiology of opioid misuse and inform the targeted deployment of public health services and interventions to prevent morbidity and mortality associated with opioid misuse.

%

\bibliographystyle{apalike}

\bibliography{references}



\section*{Supplementary material}

If $n^{(a)}$ is large, as in our application, an atom-level spatio-temporal random effect makes the model computationally expensive. We assume $f_t^{(o)}(B_{\ell j}) = \mathbf{B}(B_{\ell j})'\boldsymbol{\omega}_t^{(o)}$, where $\mathbf{B}(B_{\ell j})$ is a $q$-dimensional vector of Moran's I basis functions at atom $B_{\ell j}$ (Hughes & Haran 2013). More specifically, for our application the basis functions are the $q$ eigenvectors of the Moran operator,

$$\left(\mathrm{I} - \mathbf{X}_1^{(o)} \left(\mathbf{X}_1^{(o)\prime}\mathbf{X}_1^{(o)}\right)^{-1} \mathbf{X}_1^{(o)\prime}\right) \mathbf{A} \left(\mathrm{I} - \mathbf{X}_1^{(o)} \left(\mathbf{X}_1^{(o)\prime}\mathbf{X}_1^{(o)}\right)^{-1} \mathbf{X}_1^{(o)\prime}\right),$$

corresponding to the $q$ largest, positive eigenvalues, where $\mathbf{X}_1^{(o)}$ is the design matrix used in (6) in year $t = 1$ and $\mathbf{A}$ is the atom-level adjacency matrix. Note that the atom-level covariates used in our application are static, so $\mathbf{X}_t = \mathbf{X}_1$ for all $t$, but that does not have to hold generally.

The vector $\boldsymbol{\omega}_t^{(o)}$ is a $q$-dimensional vector of coefficients for the basis functions for indicator $o$ at time $t$. We account for temporal dependence by assuming an autoregressive model of order one for each of the $i = 1, ..., q$ coefficients. That is,

$$\begin{aligned} \omega_{1,i}^{(o)} &\sim N\left(0, \tau_{o,i}^2\right) \quad \text{for } t = 1 \\ \omega_{t,i}^{(o)} &\sim N\left(\phi_i^{(o)}\omega_{t-1,i}^{(o)}, \tau_{o,i}^2\right) \quad \text{for } t \geq 1, \end{aligned}$$

where $\tau_{o,i}^2$ are variance parameters and $\phi_i^{(o)}$ are autocorrelation parameters for each basis function $i = 1, ..., q$ and each ZCTA-level outcome $o$.

# 3 Supplemental Simulation Data and Results

## 3.1 Simulated Latent Atom-Level Counts of Misuse

In this section, we show maps of the simulated latent atom-level prevalence of misuse, $N_t(A_{it})/P_t(A_{ij})$, for the 8th and 23rd simulated data sets. The atom-level counts were simulated assuming the FNCH distribution described in the main manuscript, and as such, the atoms within a county are forced to sum to the county-level value $N_{it}$, which is assumed to be the same for all 100 simulate data sets. Figure 8 shows maps for the entire state of Ohio. To more easily visualize the spatial heterogeneity, we show the same maps but zoomed in to only include central Ohio counties (Figure 9) and southern Ohio counties (Figure 10). Central Ohio counties are defined to include Franklin, Delaware, Licking, Fairfield, Pickaway, Madison, and Union counties. Southern Ohio counties include Adams,

Athens, Brown, Clermont, Gallia, Highland, Hocking, Jackson, Lawrence, Meigs, Pike, Ross, Scioto, and Vinton counties.

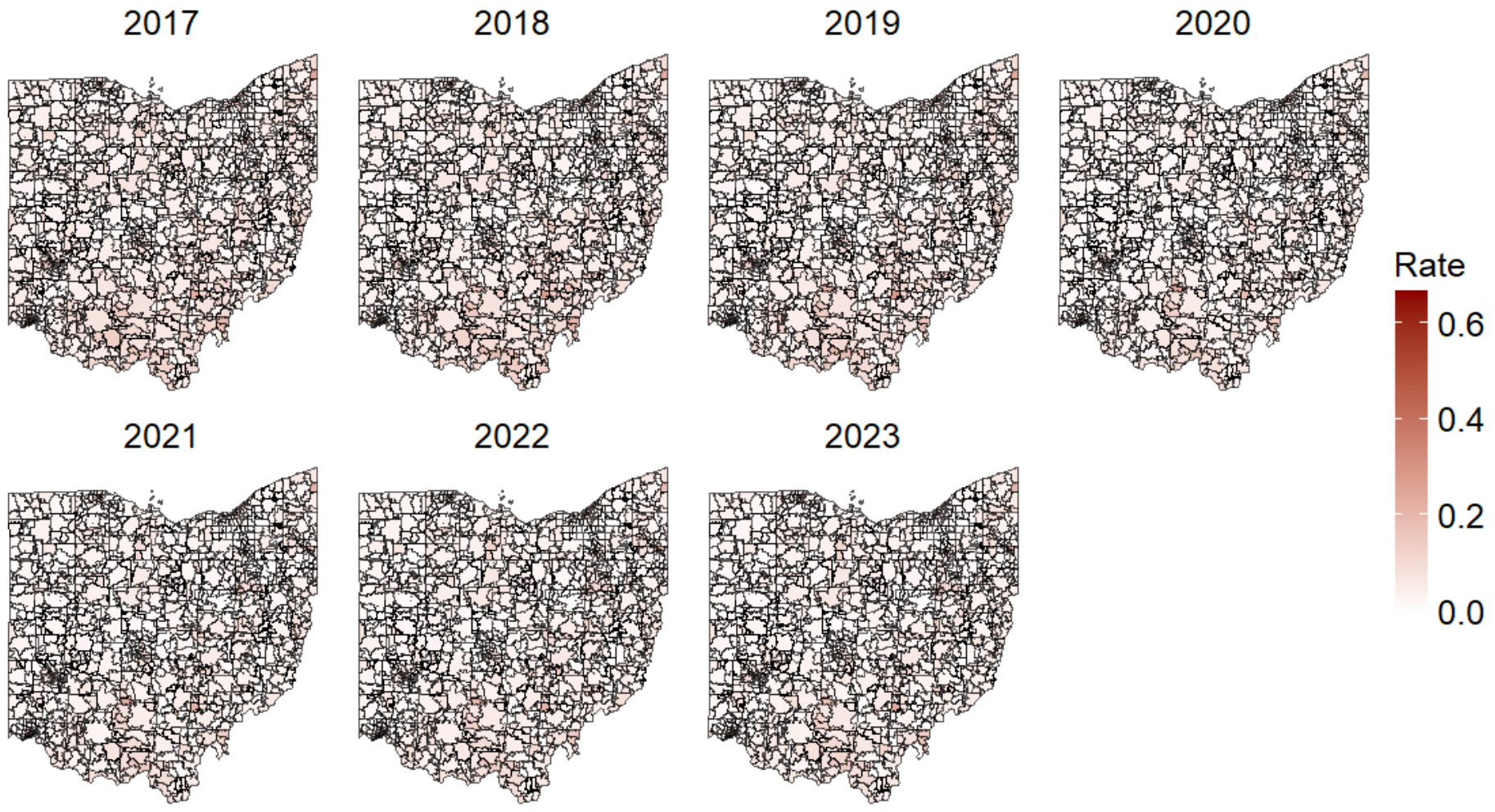


(a)

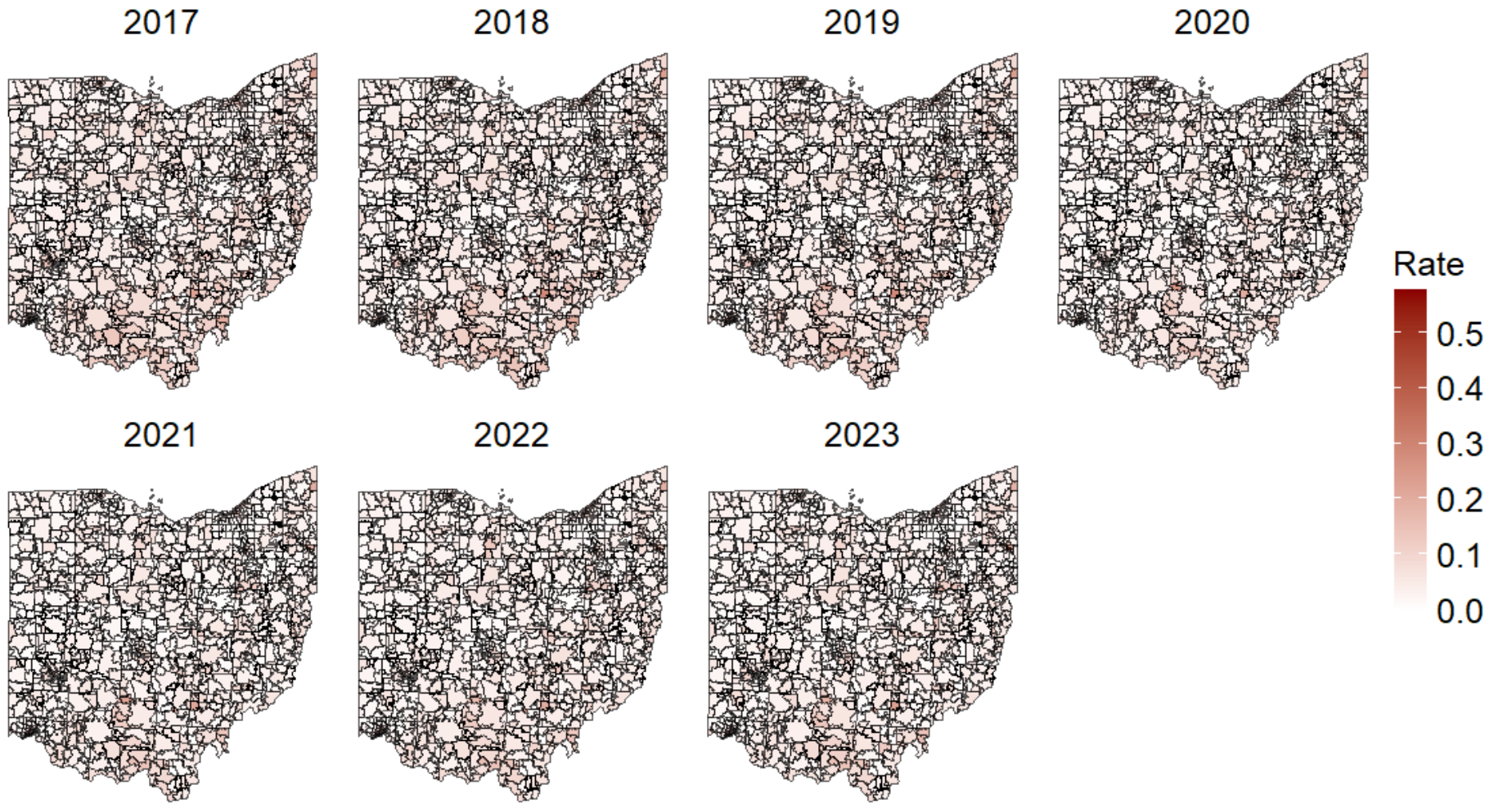


(b)

Figure 8: Simulated atom-level prevalence of misuse for the 8th (Figure a) and 23rd (Figure b) simulated data sets.

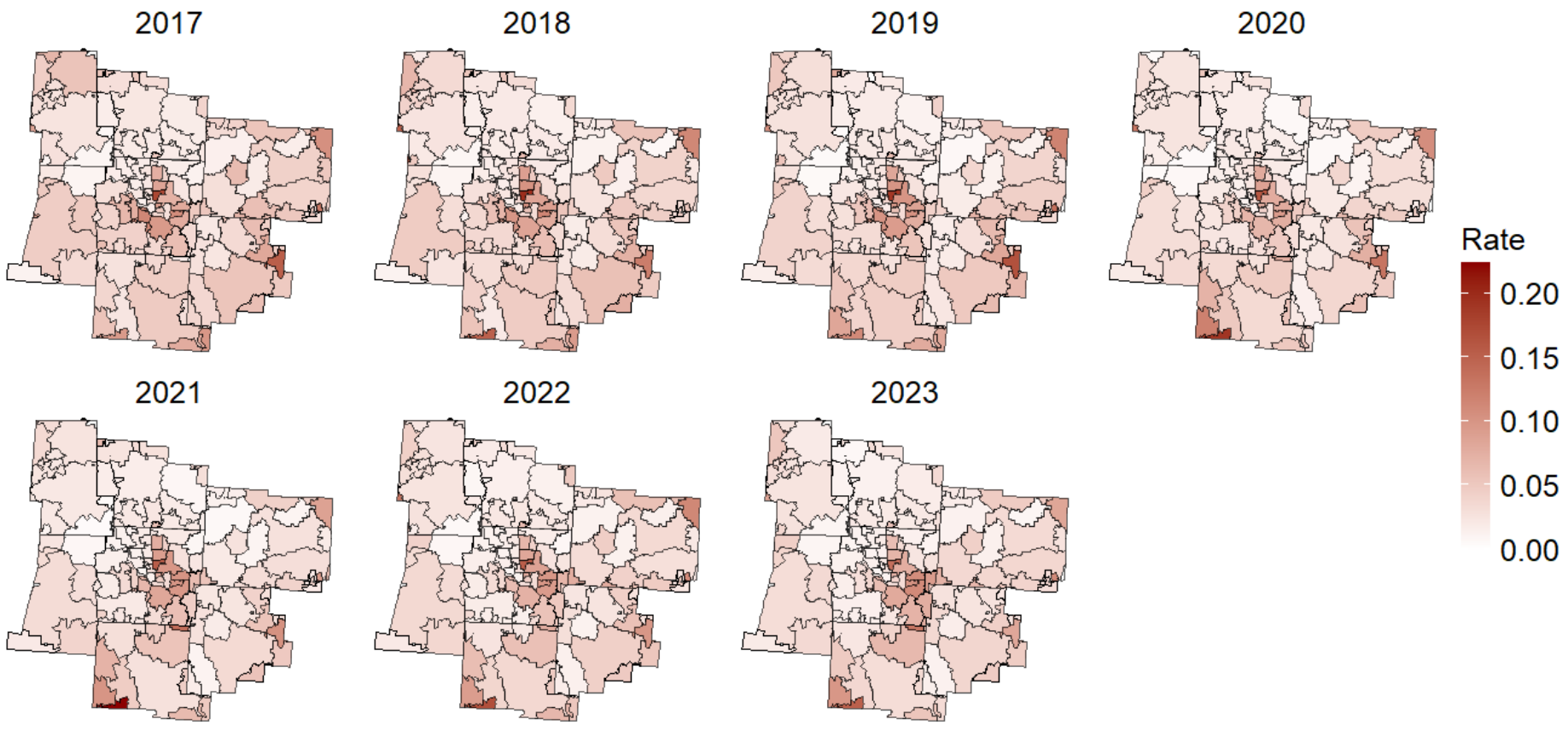


(a)

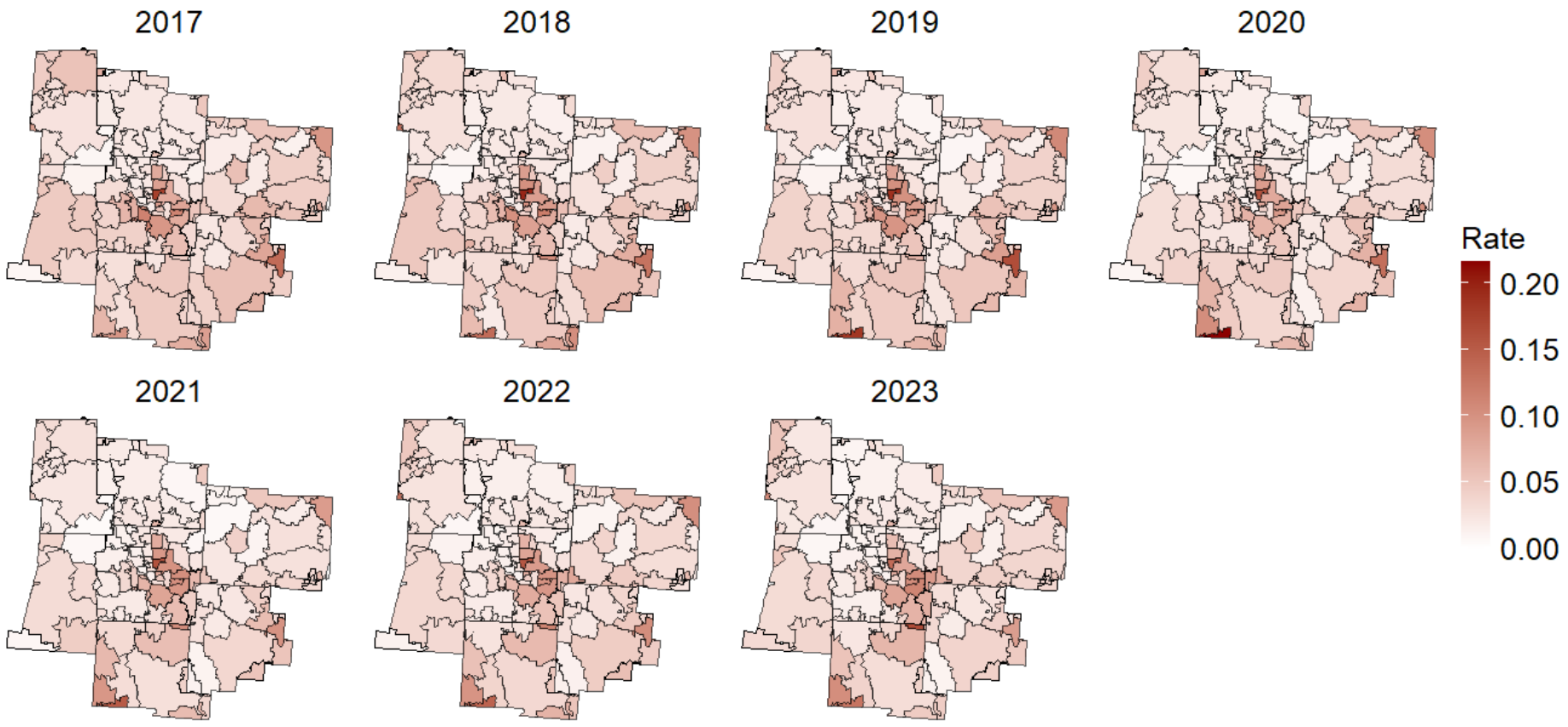


(b)

Figure 9: Simulated atom-level prevalence of misuse, zoomed into the atoms contained within central Ohio counties, for the 8th (Figure a) and 23rd (Figure b) simulated data sets.

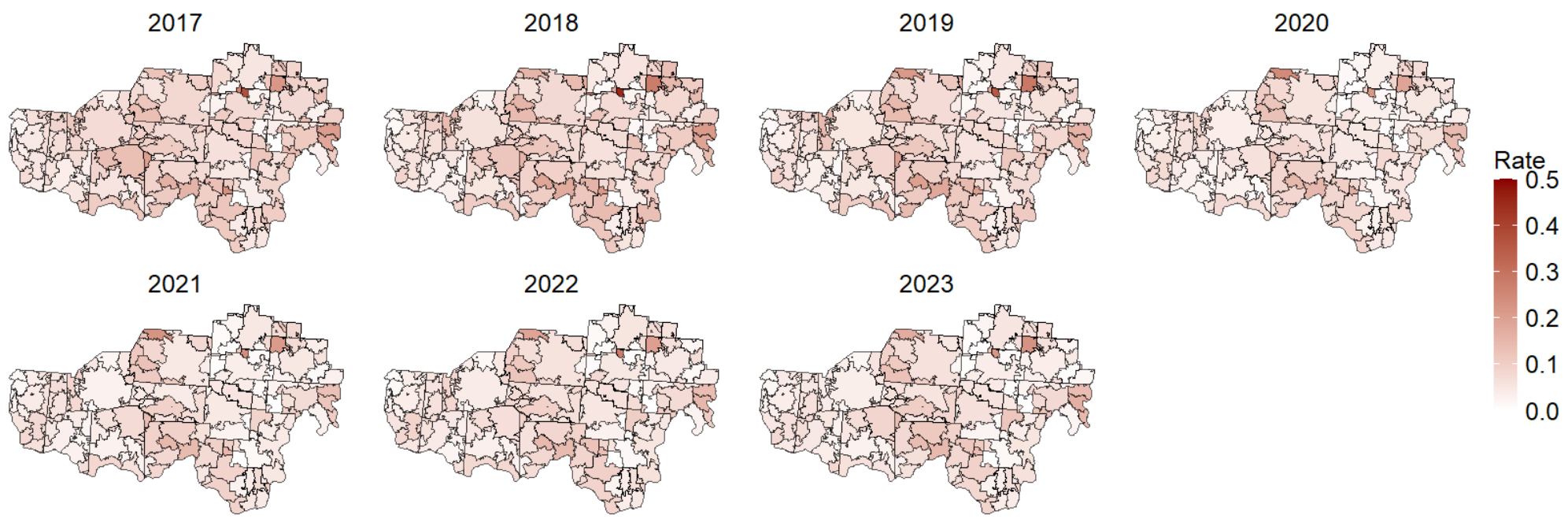


(a)

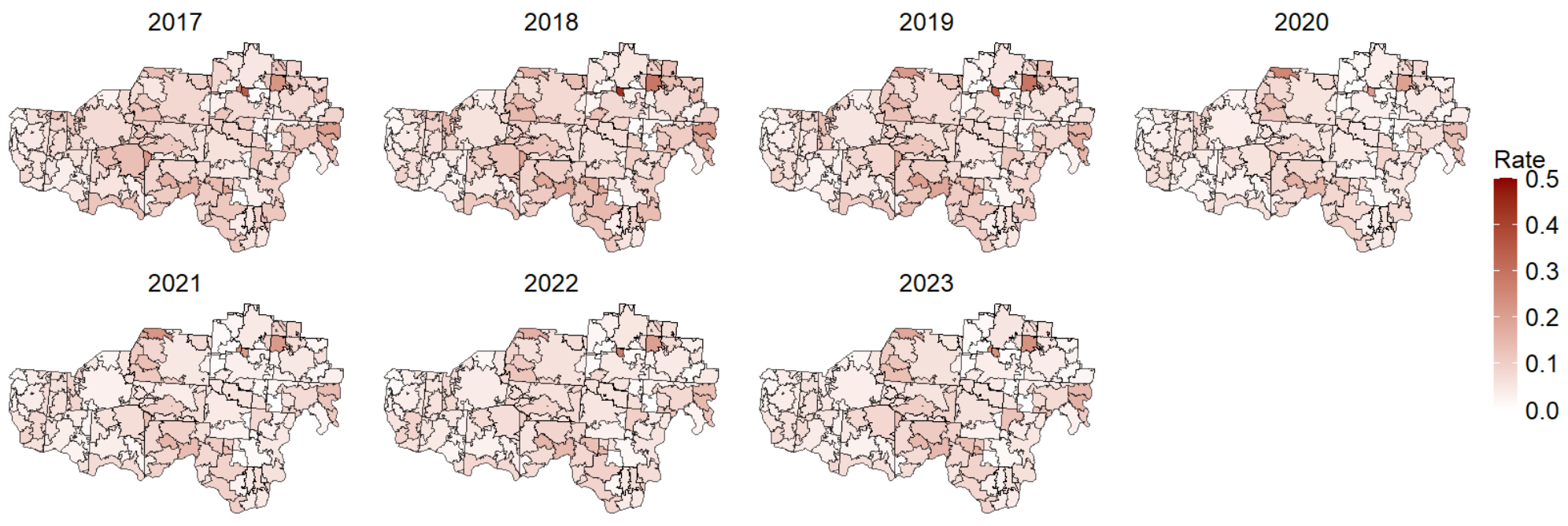


(b)

Figure 10: Simulated atom-level prevalence of misuse, zoomed into the atoms contained within southern Ohio counties, for the 8th (Figure a) and 23rd (Figure b) simulated data sets.

## 3.2 Simulated ZCTA-level Counts of Naloxone Administrations

Figures 11 - 13 show maps of the ZCTA-level rates, $Z_{\ell t}/P_{\ell t}$. The ZCTA-level counts serve as the observed "data" in the simulation study. Maps are shown for the 8th and 23rd simulated data set. Figure 11 shows the map for the entire state of Ohio; whereas Figures 12 and 13 zoom in to show just the ZCTA that intersect with central Ohio and southern Ohio counties, respectively.

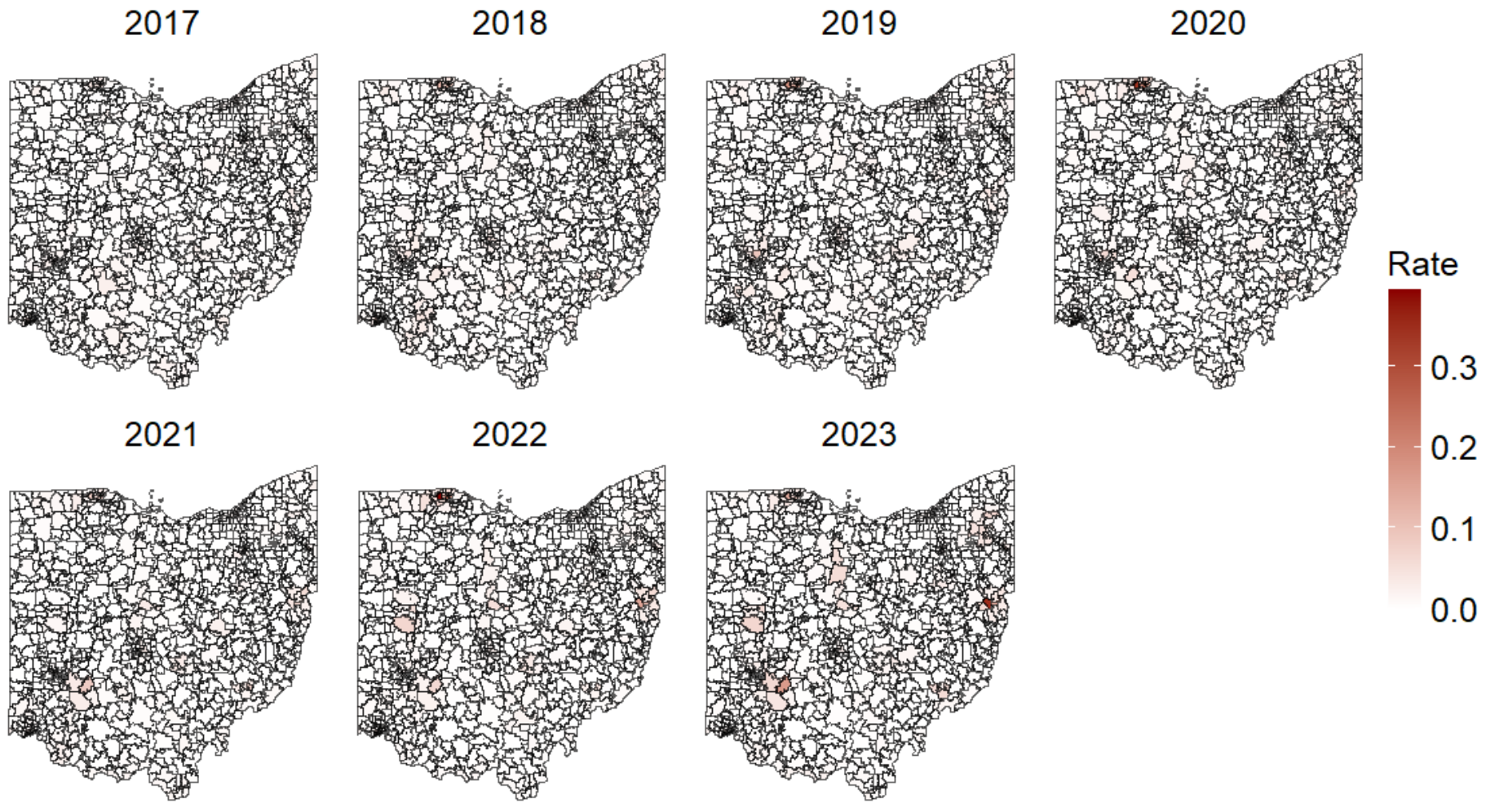


(a)

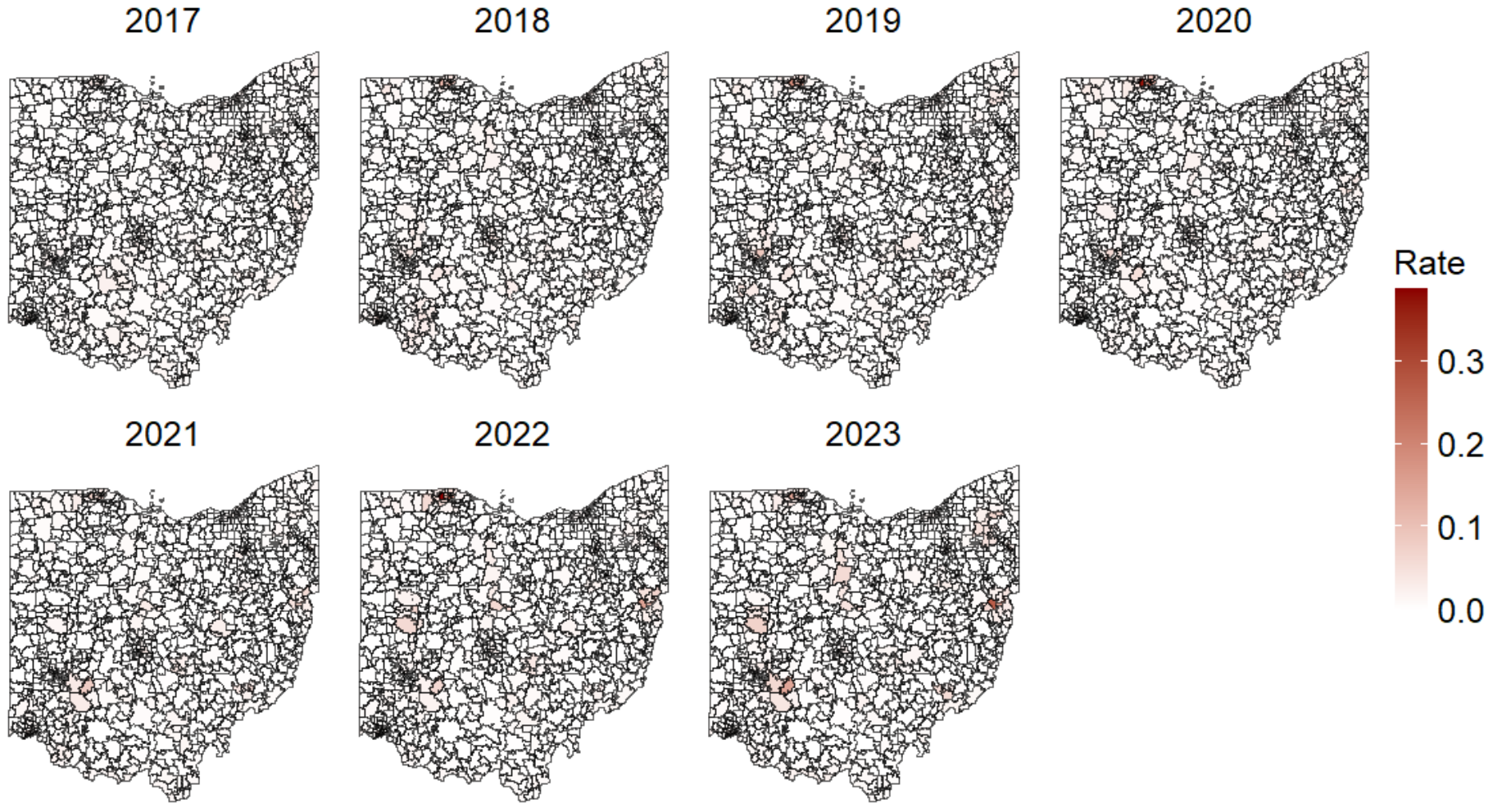


(b)

Figure 11: Simulated ZCTA-level rates of naloxone administrations for the 8th (Figure a) and 23rd (Figure b) simulated data sets.

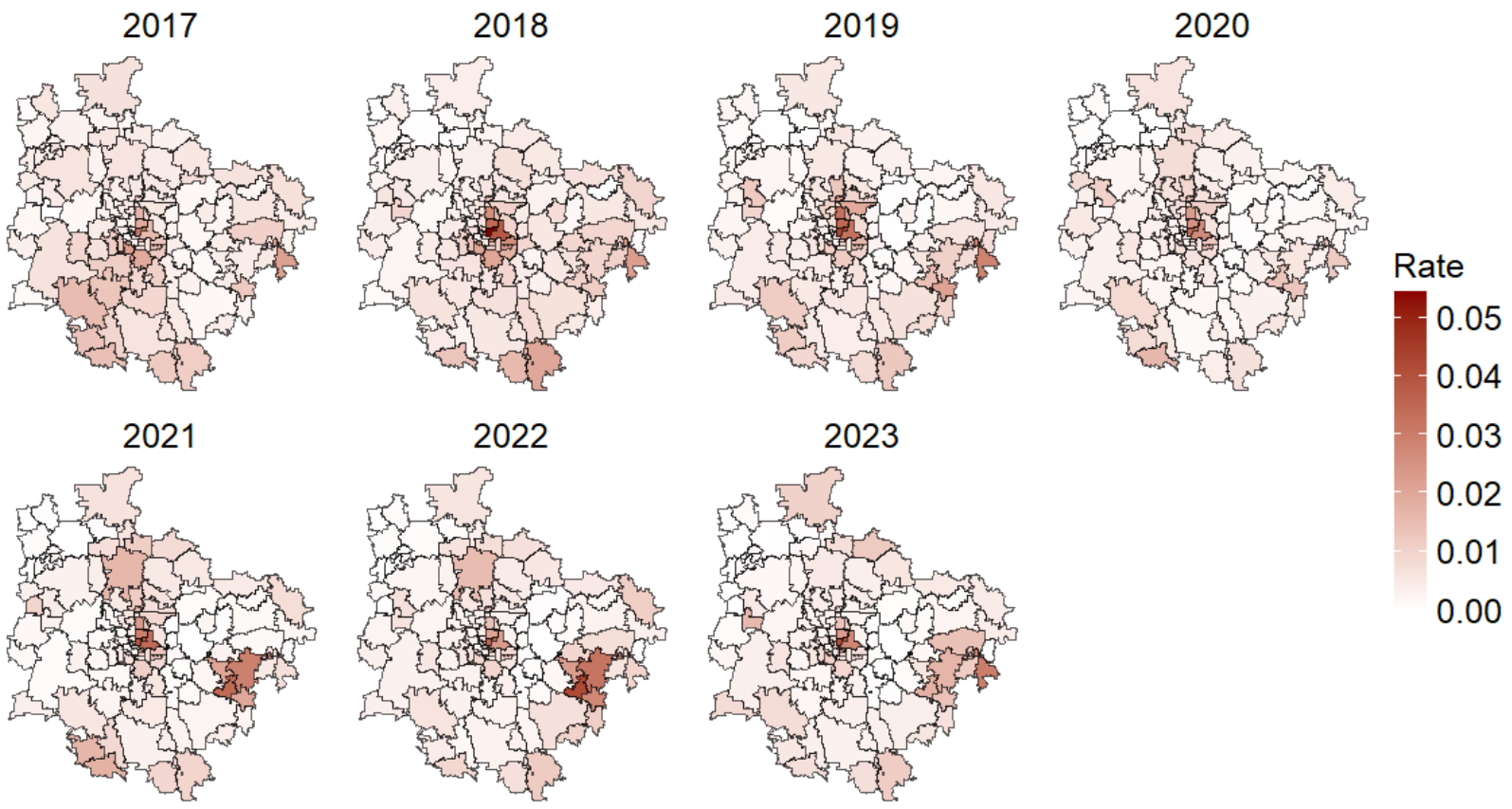


(a)

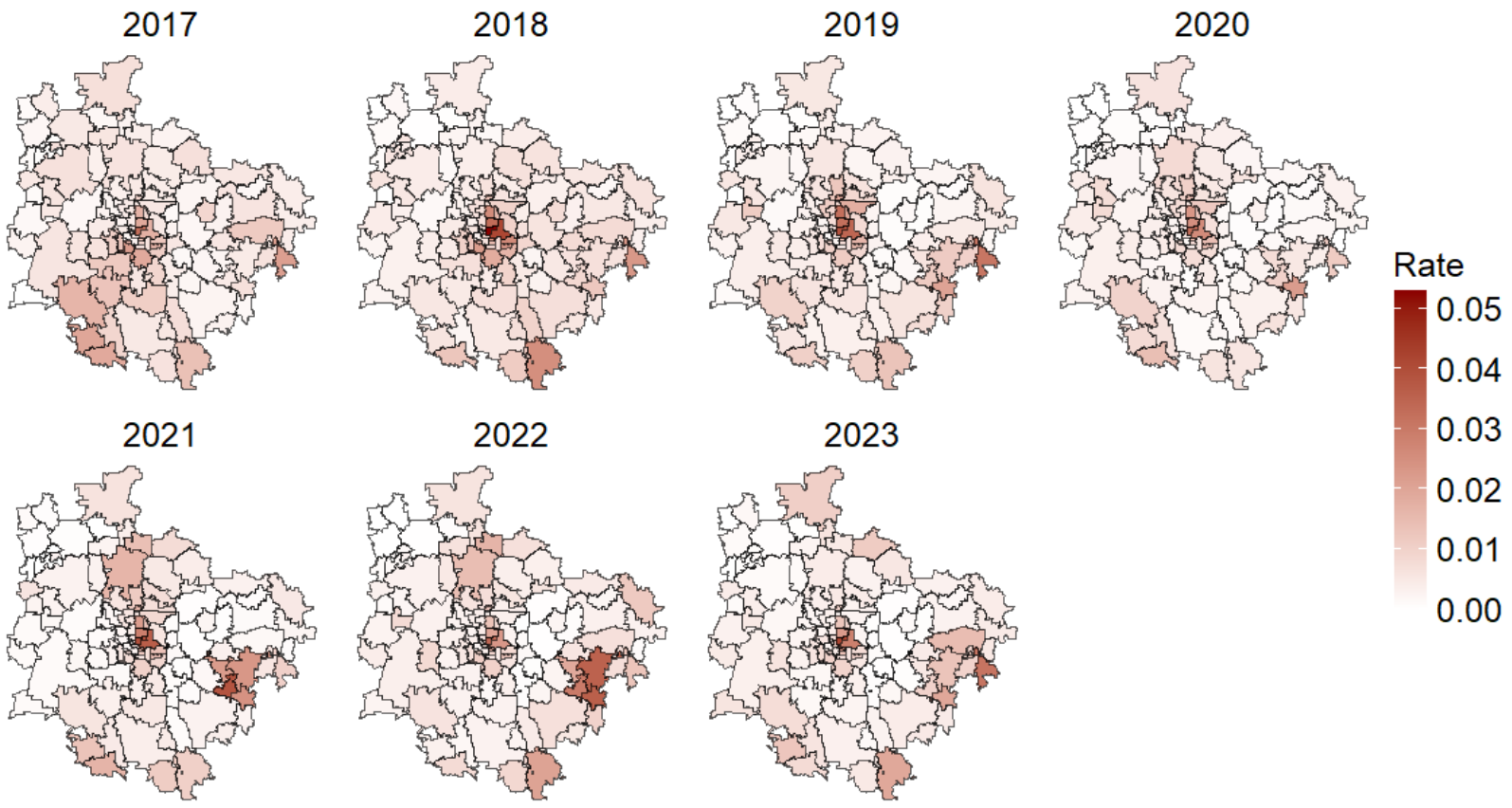


(b)

Figure 12: Simulated ZCTA-level rates of naloxone administrations, zoomed to show the ZCTA that intersect central Ohio counties, for the 8th (Figure a) and 23rd (Figure b) simulated data sets.

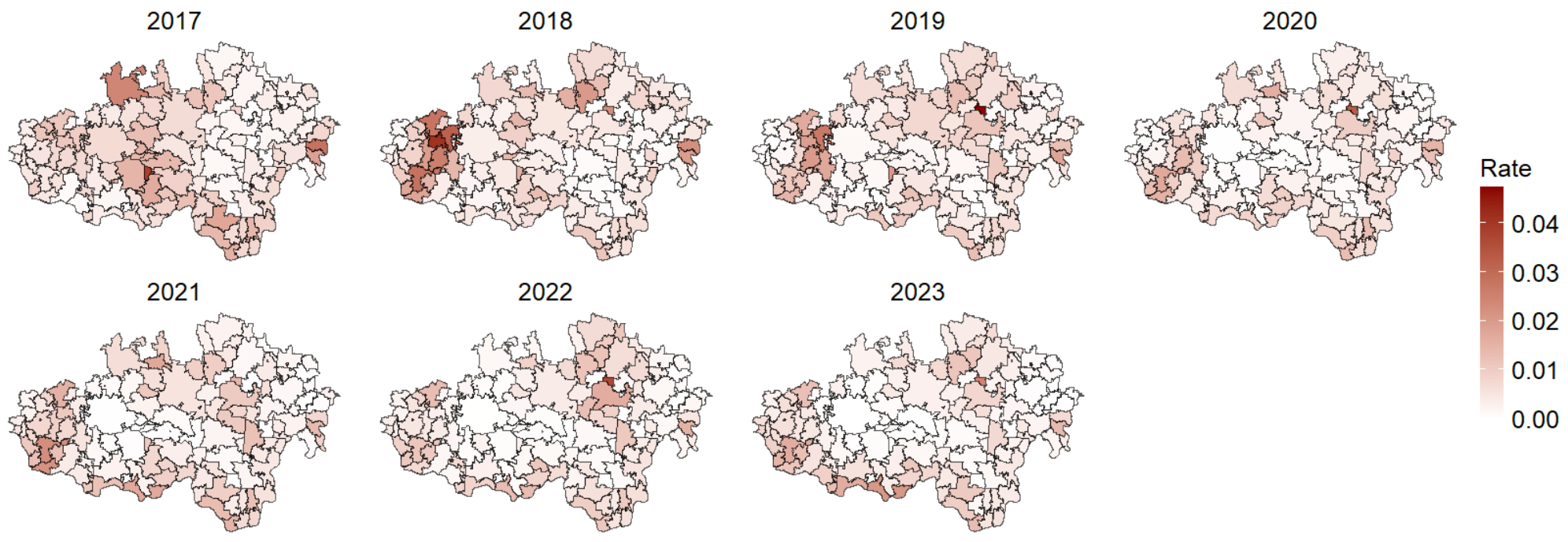


(a)

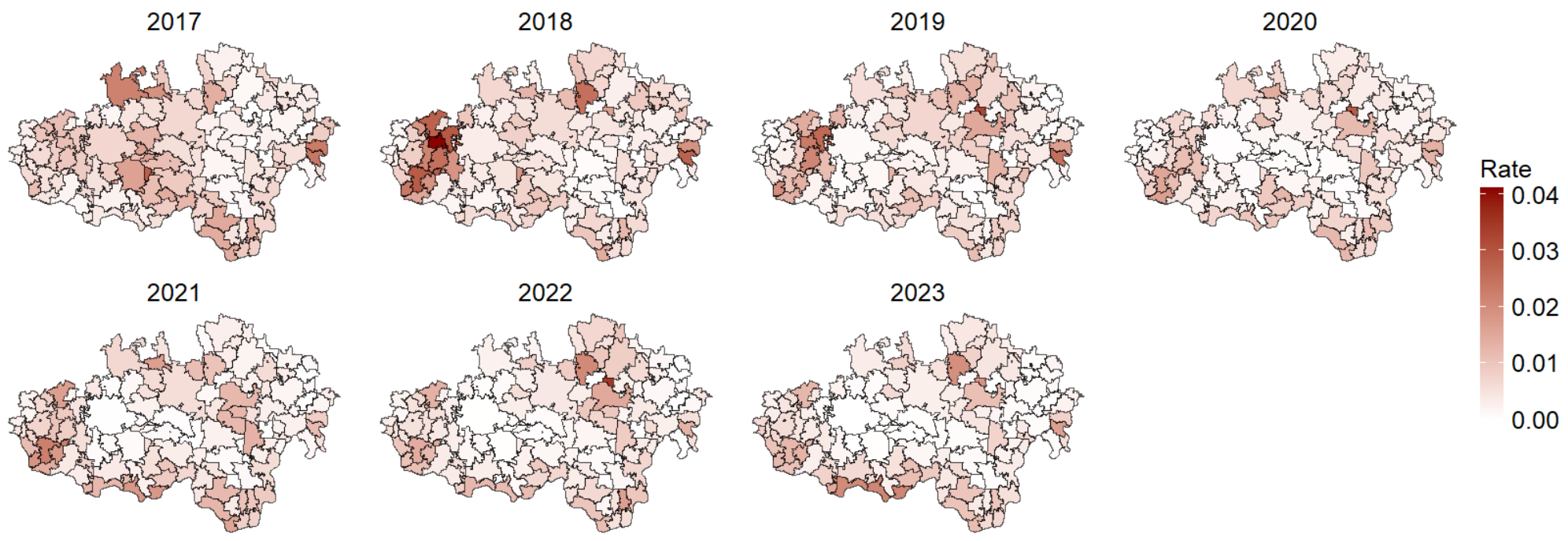


(b)

Figure 13: Simulated ZCTA-level rates of naloxone administrations, zoomed to show the ZCTA that intersect southern Ohio counties, for the 8th (Figure a) and 23rd (Figure b) simulated data sets.

### 3.3 Maps of Estimated Atom-Level Rates of Misuse

This section shows the corresponding maps of the estimated atom-level prevalence of misuse, $\hat{N}_t(A_{ij})/P_t(A_{ij})$, resulting from the proposed MIA model. The results shown here are the result of fitting the proposed model to the 8th and 23rd simulated data sets shown in the previous section, and the corresponding truth are the maps in Section 3.1.

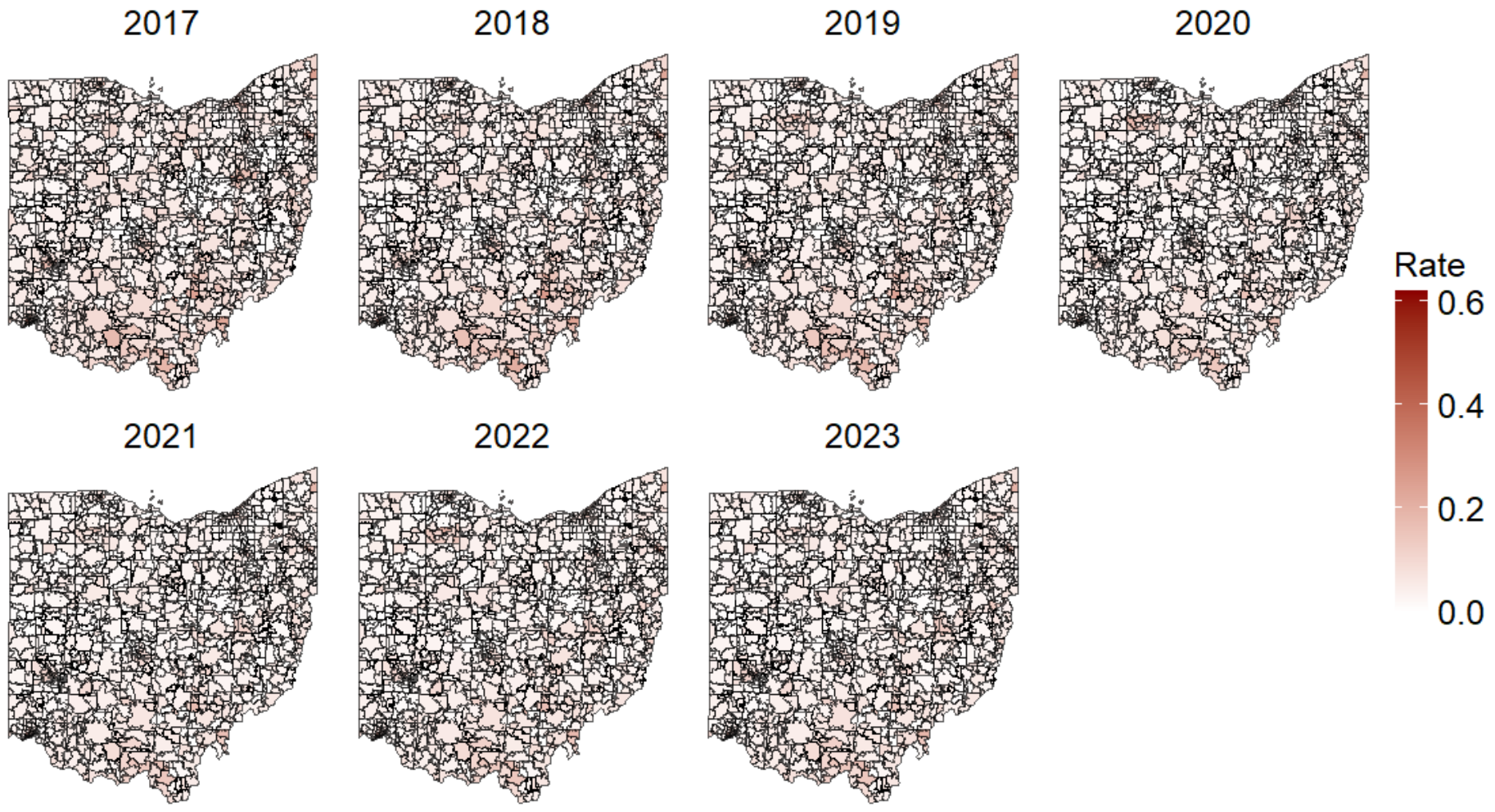


(a)

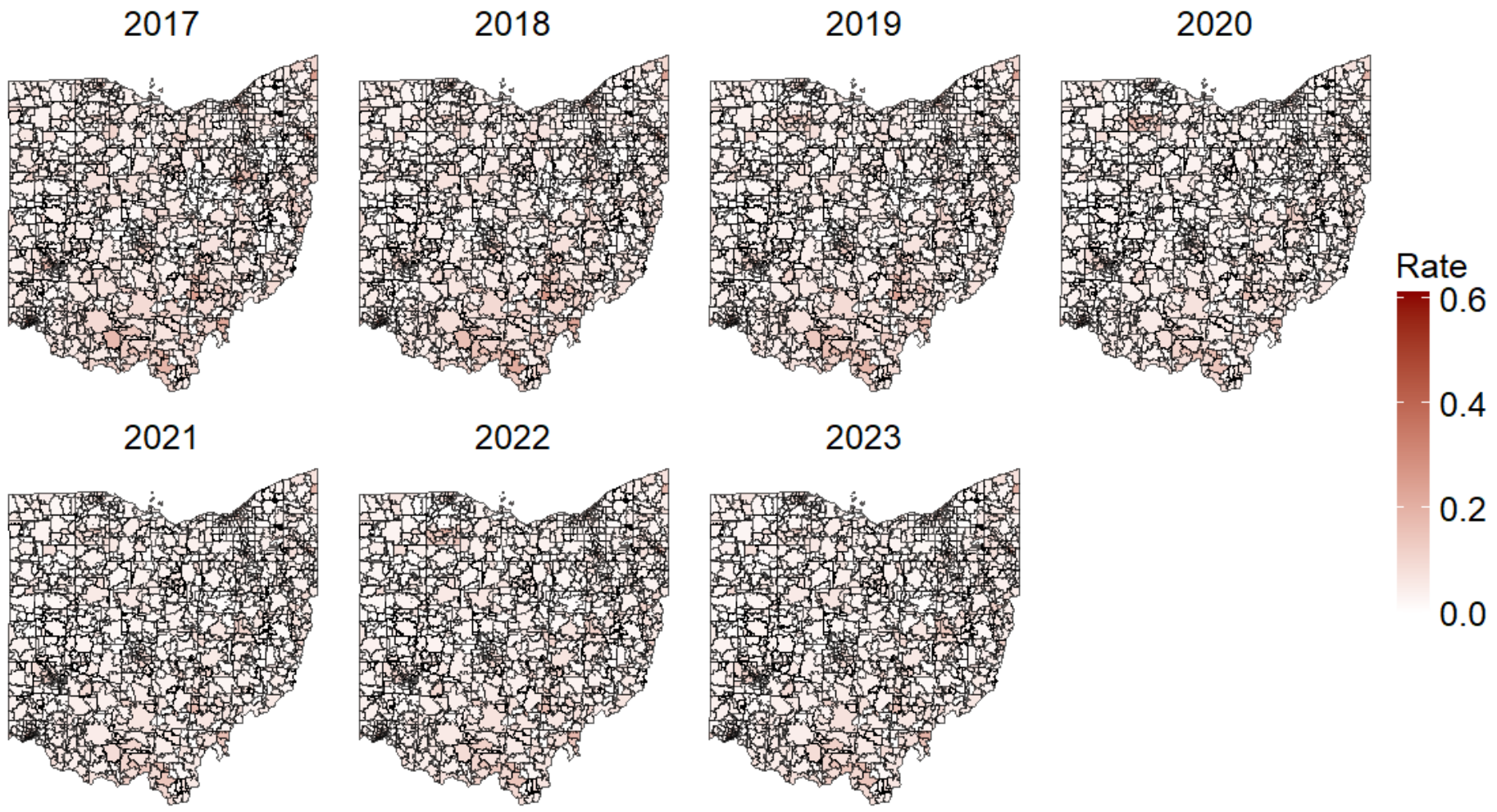


(b)

Figure 14: Estimated atom-level prevalence of misuse after fitting the 8th (Figure a) and 23rd (Figure b) simulated data sets.

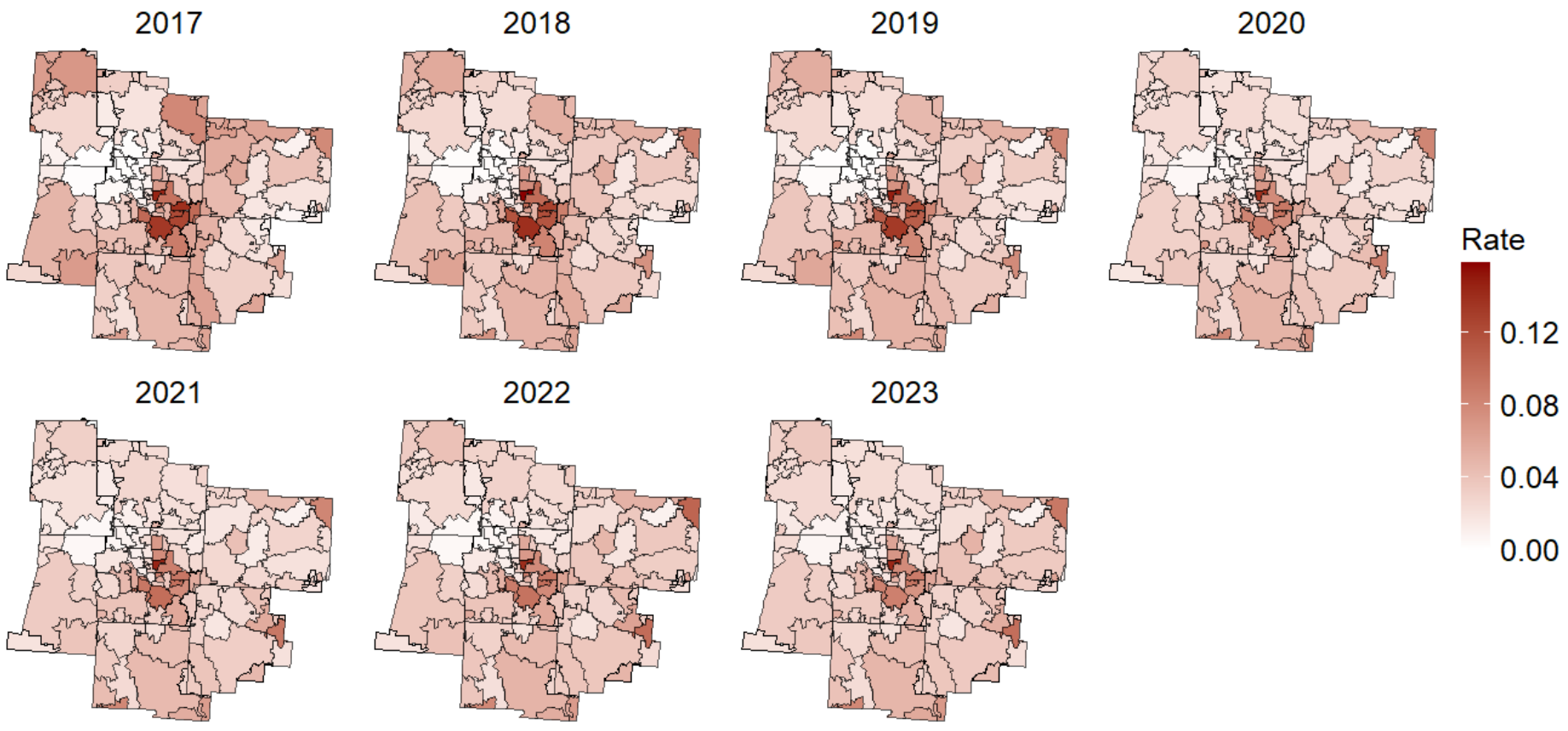


(a)

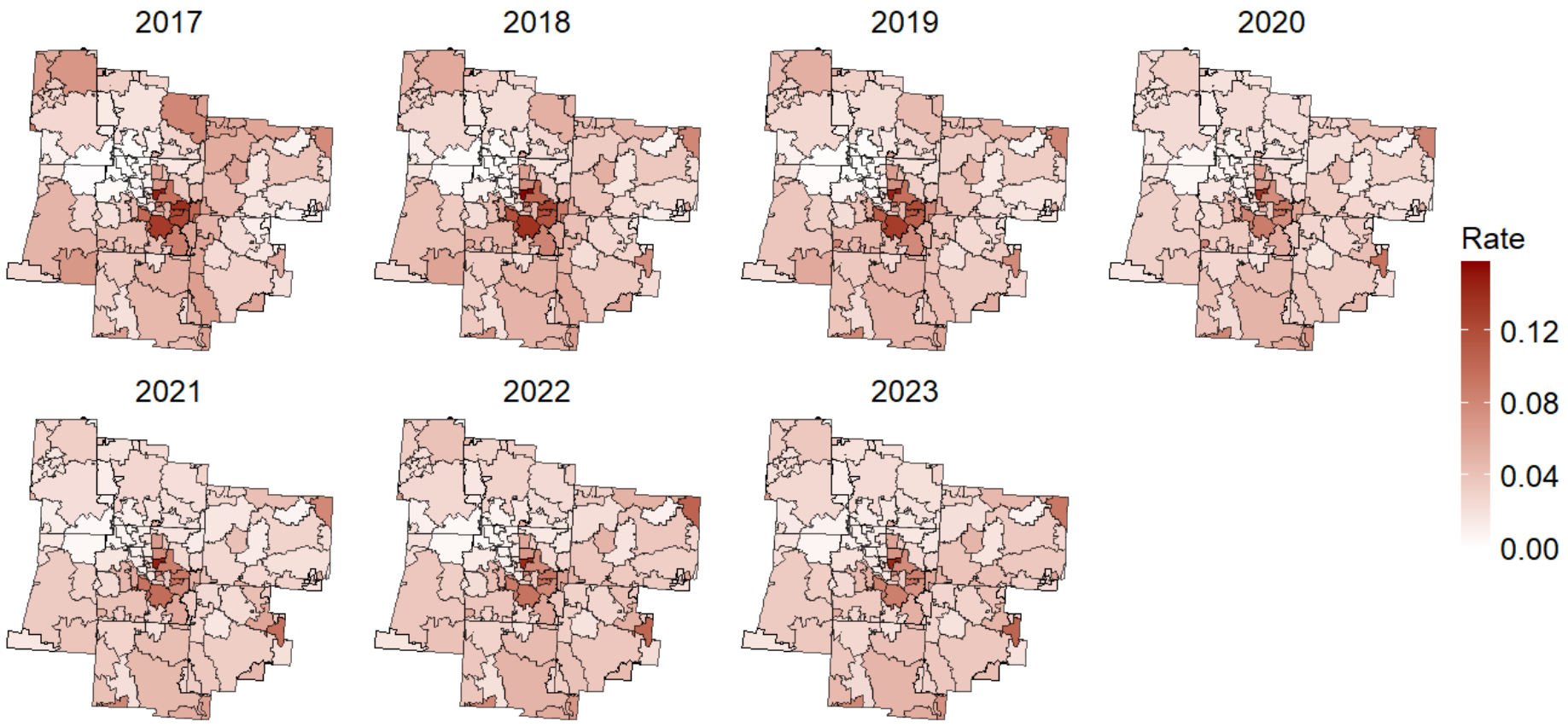


(b)

Figure 15: Estimated atom-level prevalence of misuse, zoomed into the atoms contained within central Ohio counties, after fitting the 8th (Figure a) and 23rd (Figure b) simulated data sets.

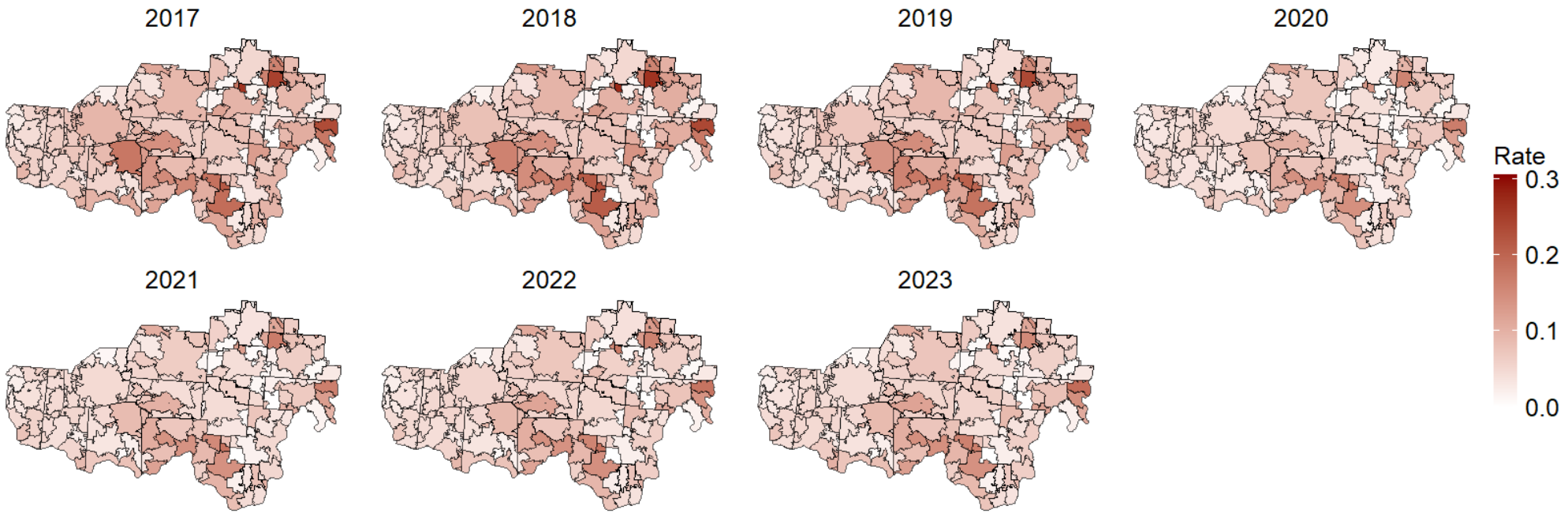


(a)

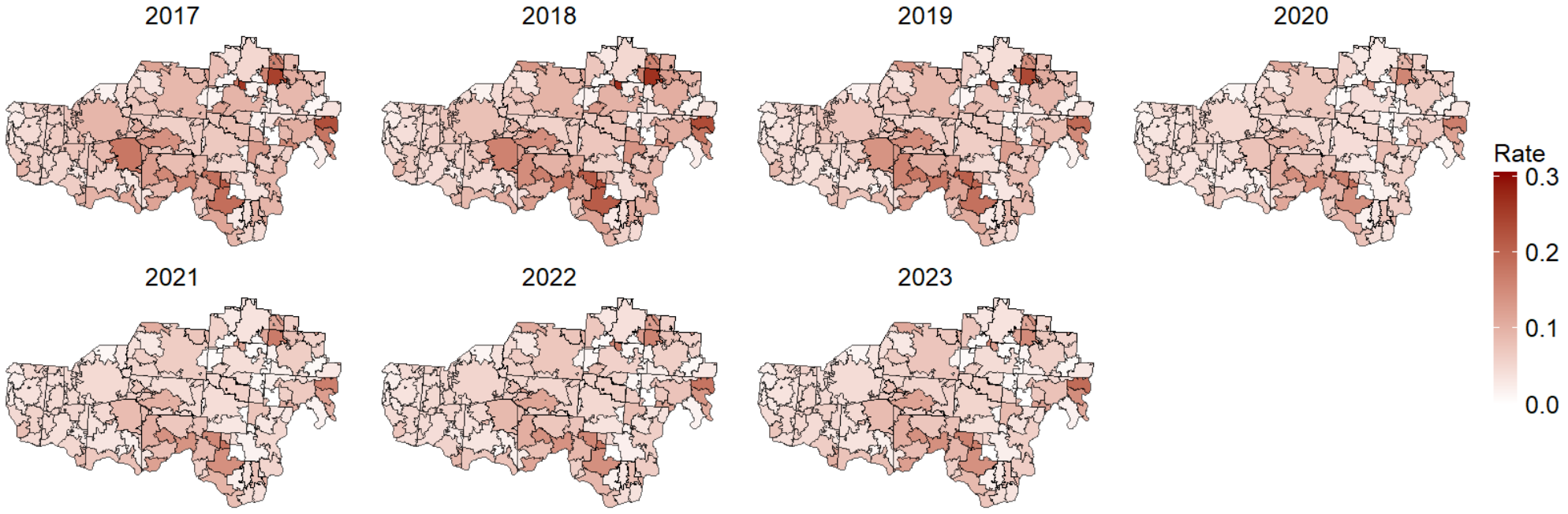


(b)

Figure 16: Estimated atom-level prevalence of misuse, zoomed into the atoms contained within southern Ohio counties, after fitting the 8th (Figure a) and 23rd (Figure b) simulated data sets.

# 4 Additional Simulation Results

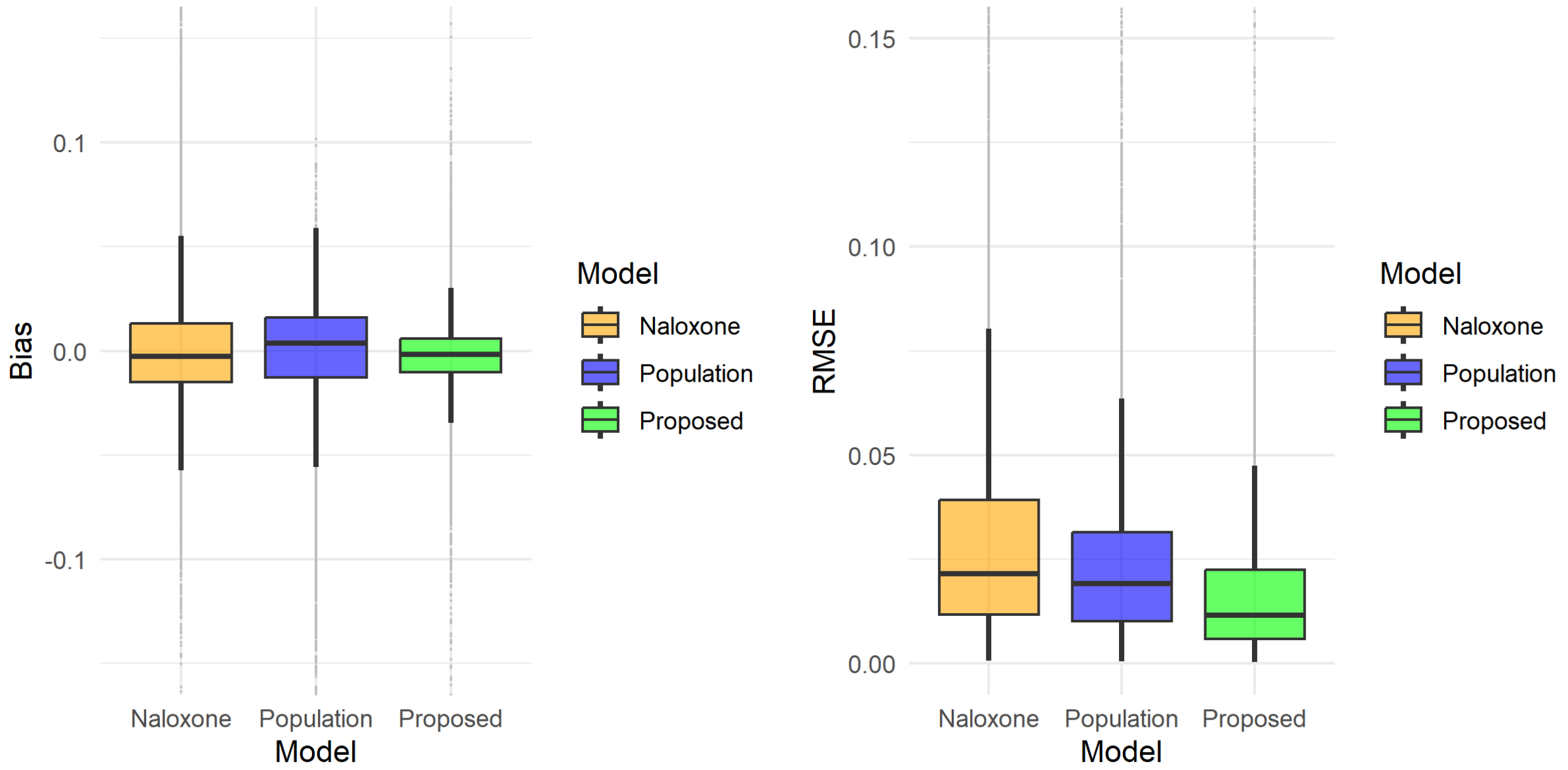


Figure 17: Boxplots of the empirical bias (left) and RMSE (right) for each location-time combination for the proposed MIA model estimates, the comparison estimates that downscale proportionally to naloxone counts, and the comparison estimates that downscale proportionally to population size.

Figure 17 contains boxplots of the empirical bias and root mean-squared error (RMSE) of the estimated OMP for our proposed MIA model and the two comparison models. The empirical bias is averaged over the 100 simulated data sets and computed as computed as $\text{bias}_{ijt} = \frac{1}{100}\sum_{m=1}^{100}\left(\hat{N}_t(A_{ij})/P_t(A_{ij}) - N_t(A_{ij})/P_t(A_{ij})\right)$, and the RMSE is given by $\text{RMSE}_{ijt} = \sqrt{\frac{1}{100}\sum_{m=1}^{100}\left(\hat{N}_t(A_{ij})/P_t(A_{ij}) - N_t(A_{ij})/P_t(A_{ij})\right)^2}$. For the proposed model, the estimated atom-level counts are computed as the median of the posterior distribution, and the boxplots summarize these quantities across all locations and times. Note that outliers are portrayed as light gray dots, and the figure has been zoomed in such that not all outliers appear.

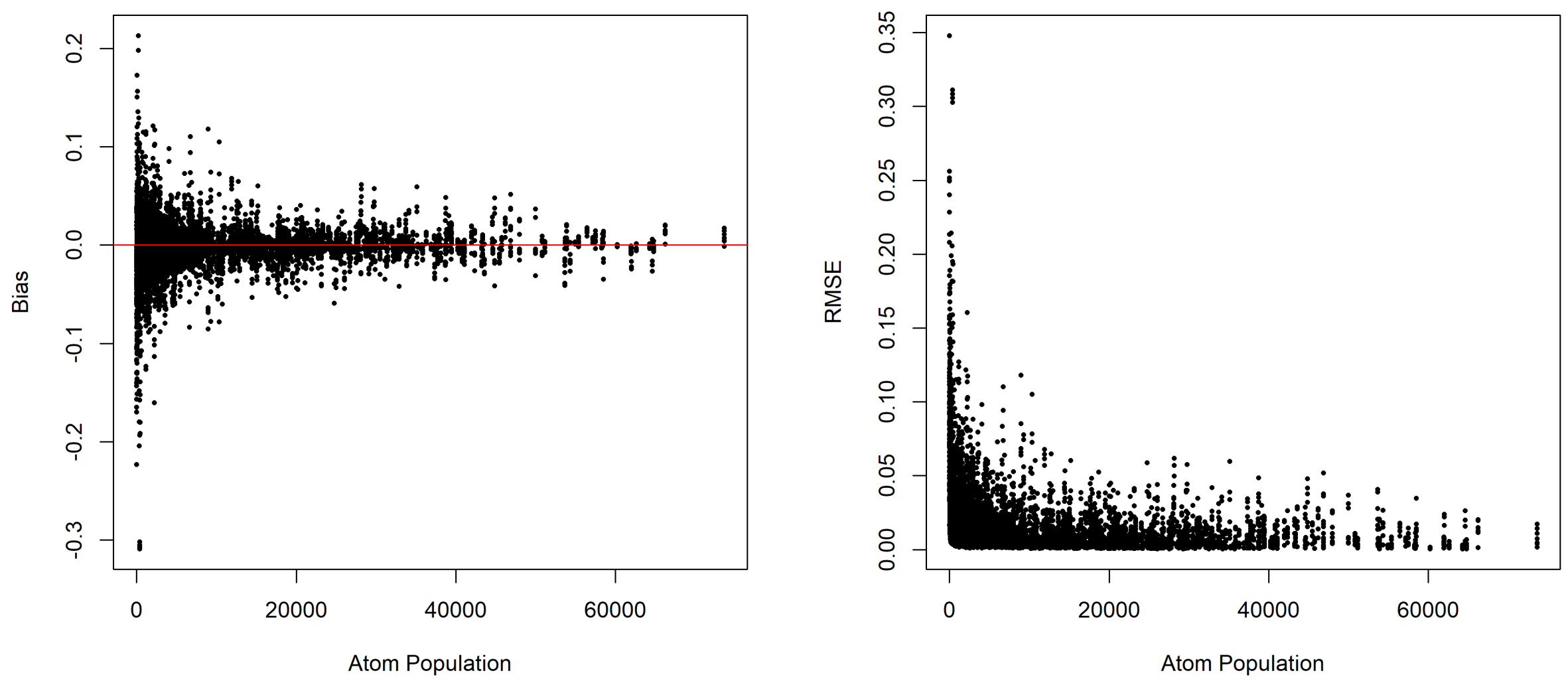


Figure 18: Plot of the empirical bias and RMSE versus the atom-level population. In general, we see the error decreases as the population increases.

# 5 Supplemental Application Results

## 5.1 County-Level Model Results

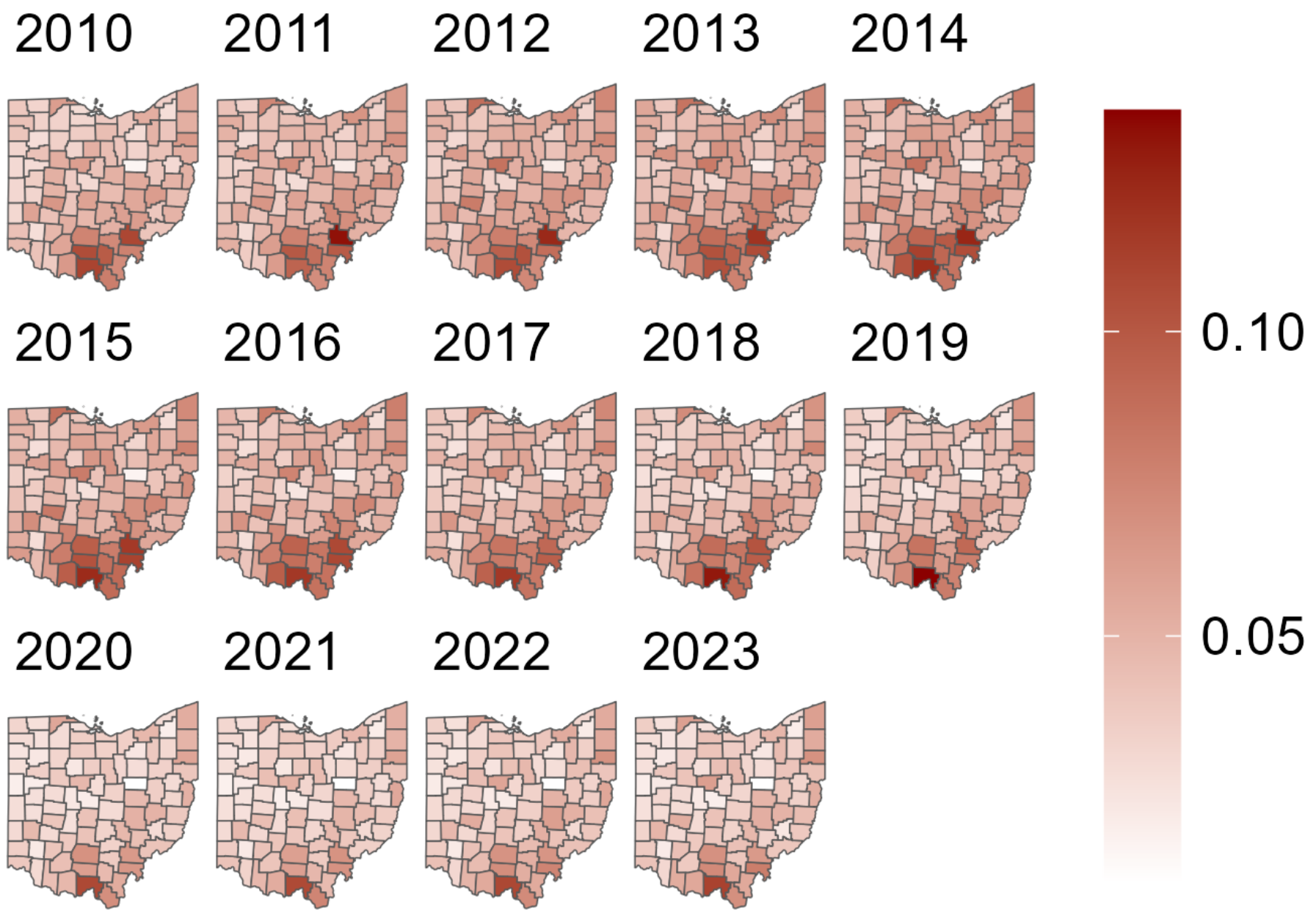


Figure 19: County-level estimated OMP for 2010 to 2023.

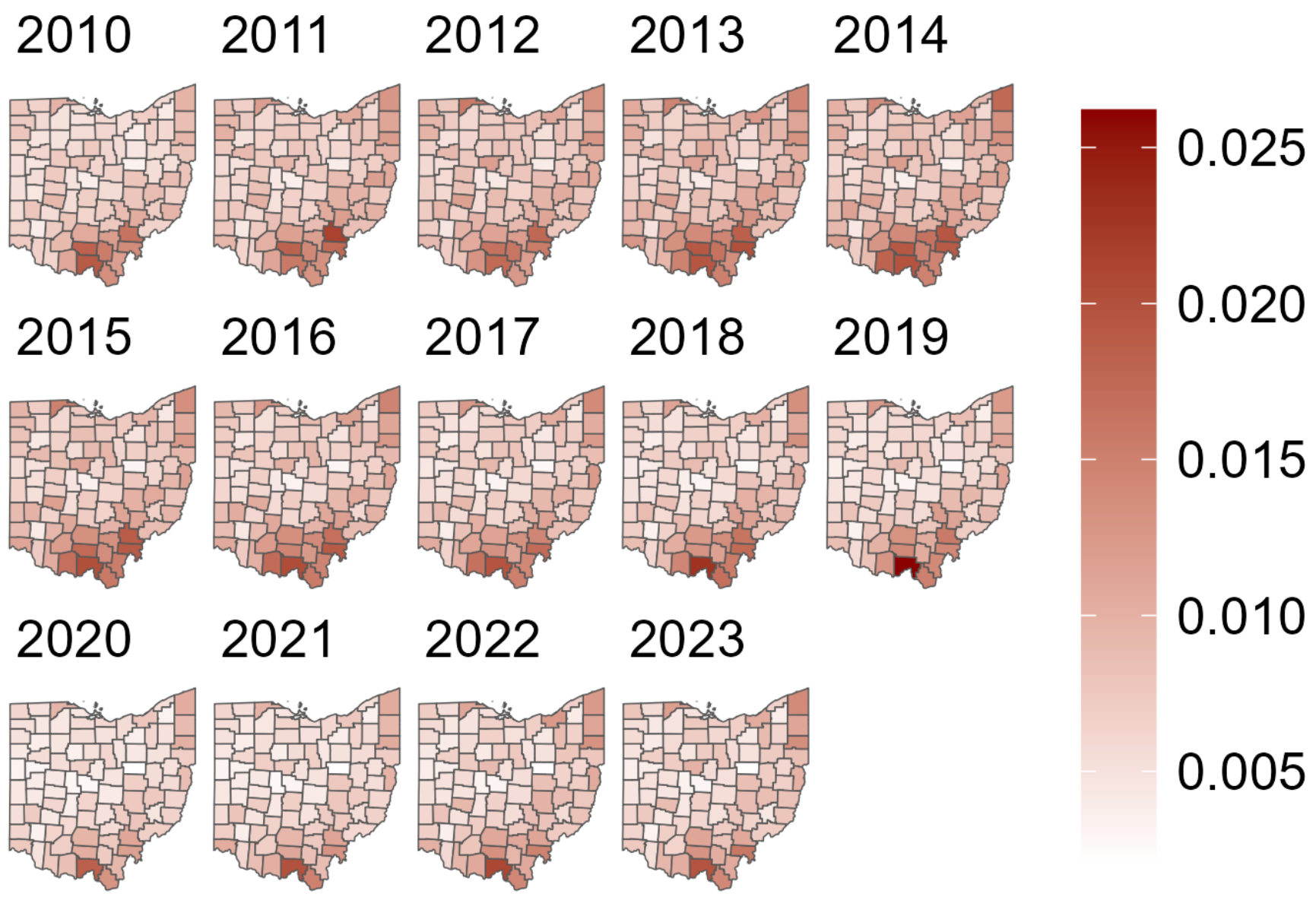


Figure 20: Standard deviation of county-level OMP for 2010 to 2023.

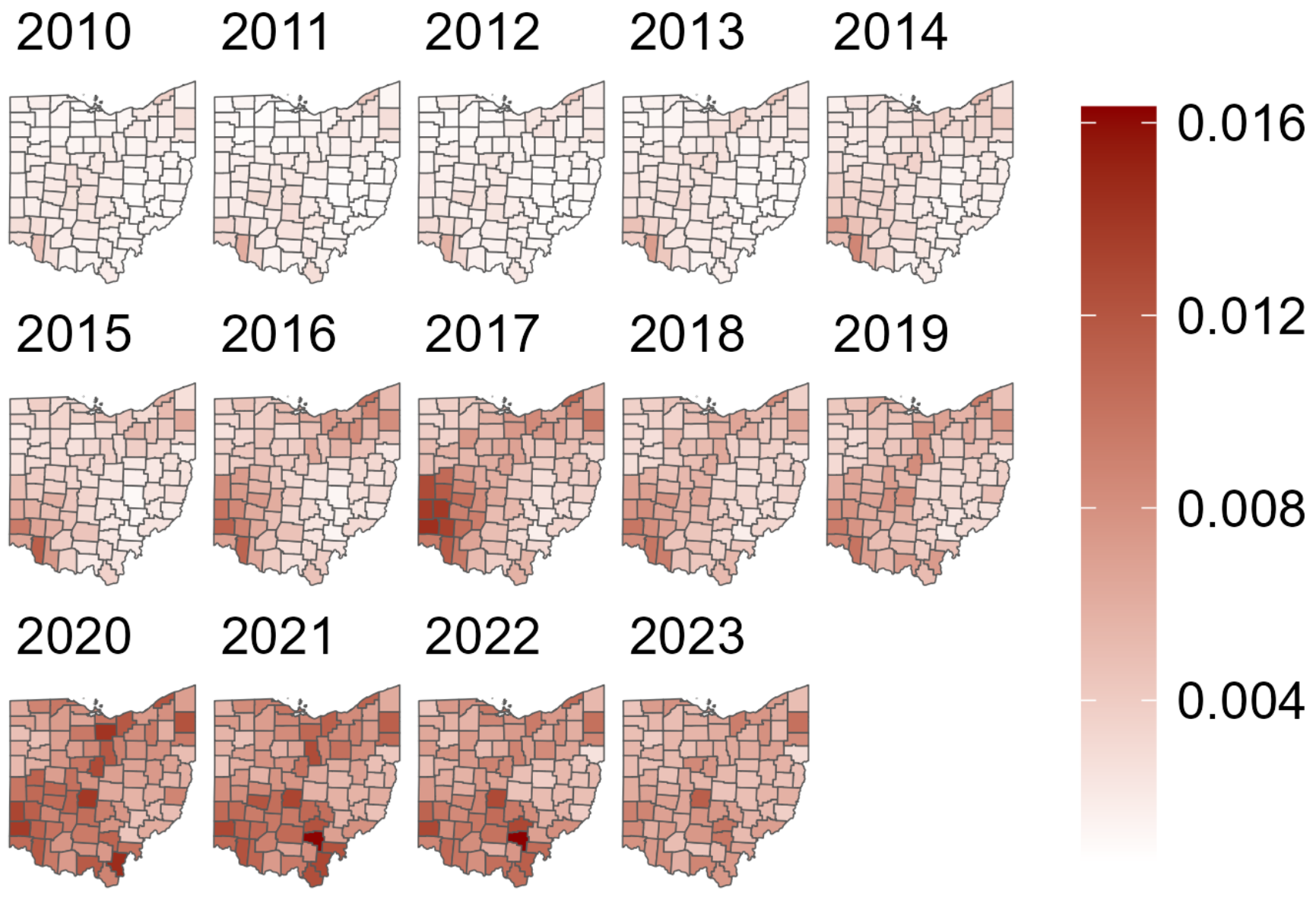


Figure 21: Estimated county-level rate of opioid overdose death among PWMO.

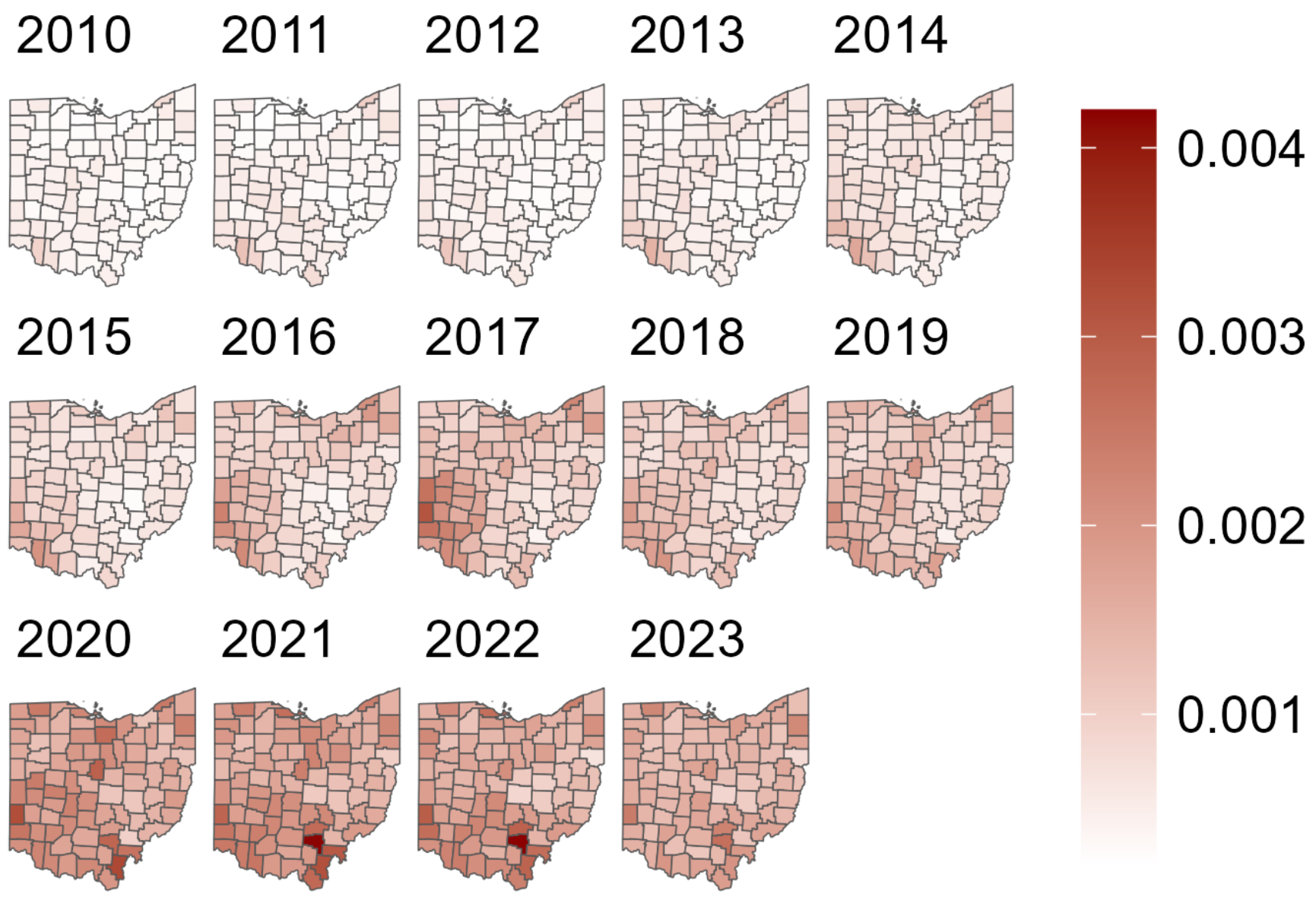


Figure 22: Standard deviation of the rate of opioid overdose death among PWMO.

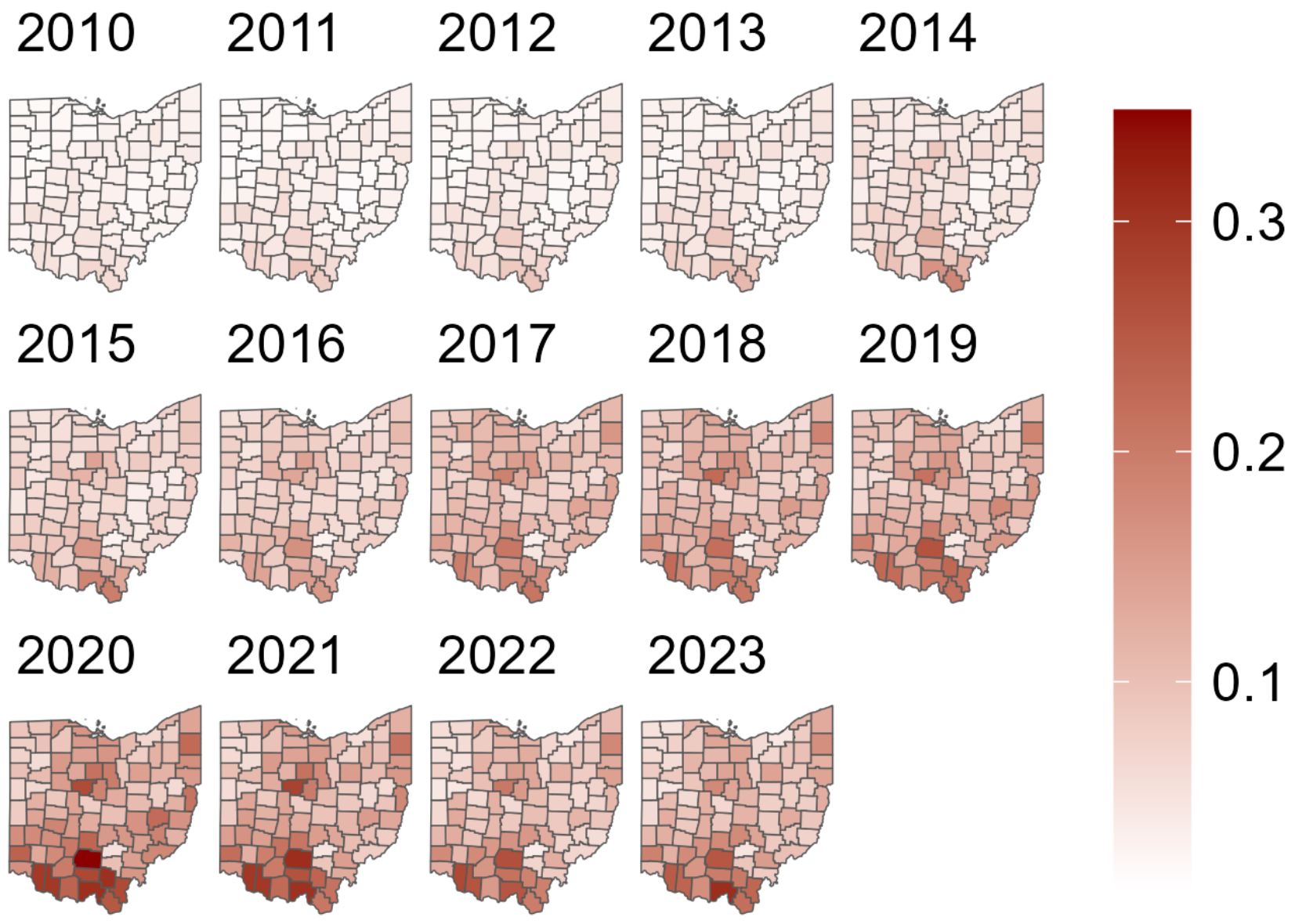


Figure 23: Estimated county-level rate of opioid treatment admissions among PWMO.

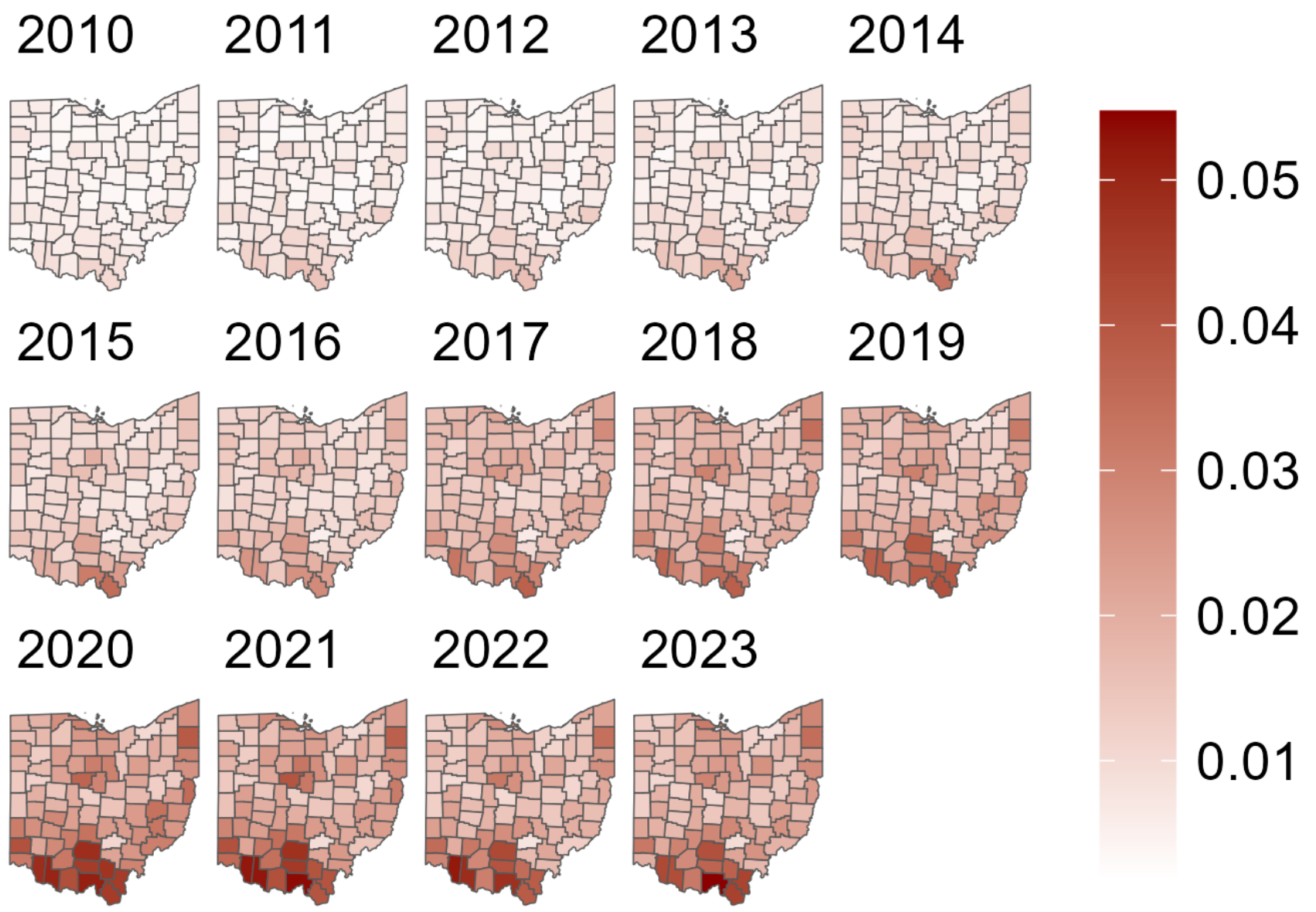


Figure 24: Standard deviation of the rate of opioid treatment admissions among PWMO.

| | Parameter | Mean | 95% CI |
|---|---|---|---|
| Treatment Rate | HPSA | -0.02 | (-0.08, 0.04) |
| | MU | 0.04 | (-0.03, 0.1) |
| Death Rate | Interstates | 0.05 | (-0.03, 0.13) |
| | HIDTA | 0.02 | (-0.06, 0.11) |
| | MSA | 0.19 | (0.1, 0.27) |
| Misuse Prevalence | Poverty | 0.18 | (0.13, 0.22) |
| | Unemployed | 0.02 | (-0.02, 0.07) |
| | High School Degree | 0.13 | (0.11, 0.17) |
| | Food Stamps | 0.31 | (0.27, 0.36) |
| | Prescribing Rate | -0.05 | (-0.08, -0.02) |

Table 2: Estimates and 95% credible intervals for the regression coefficients in the outcome-level models for the rates of treatment and death among those who misuse opioids, as well as the covariate effects of the county-level prevalence of misuse.

## 5.2 Sub-County Level Model Results

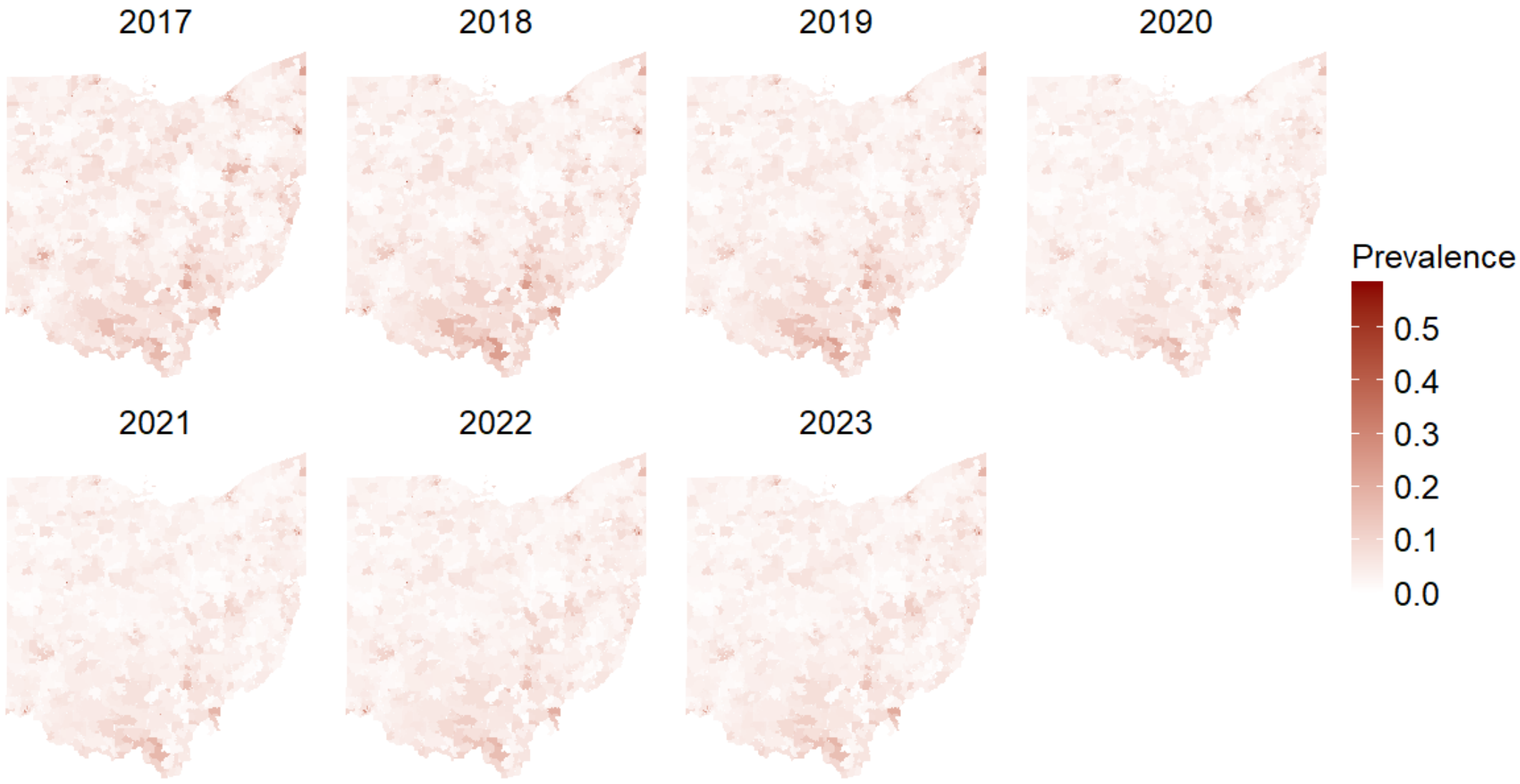


Figure 25: Estimated atom-level OMP.

| Parameter | Mean | CI |
|---|---|---|
| Intercept | -3.08 | (-3.14, -3.03) |
| Metropolitan | 0.57 | (0.5, 0.63) |
| Micropolitan | 0.01 | (-0.03, 0.05) |

Table 3: Estimates and 95% credible intervals for the regression coefficients in the ZCTA-level model of the rate of naloxone administrations among people who misuse opioids.

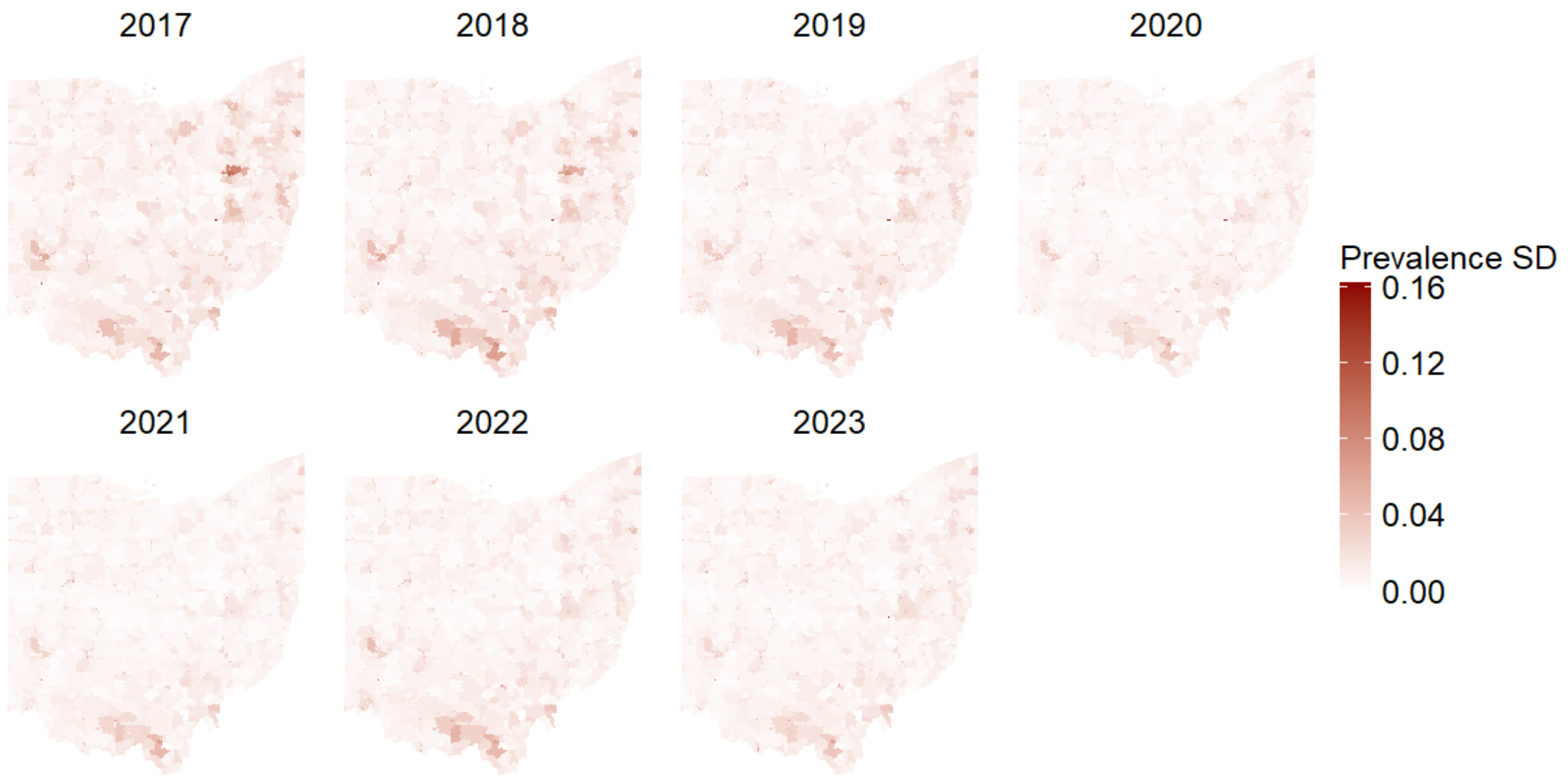


Figure 26: Posterior standard deviation of atom-level OMP.

| Parameter | Mean | 95% CI |
|---|---|---|
| Intercept | -0.77 | (-1.01, -0.49) |
| Unemployment | 0.98 | (0.9, 1.06) |
| College Degree | -0.28 | (-0.34, -0.22) |

Table 4: Estimates and 95% credible intervals for the regression coefficients in the atom-level model of prevalence of people who misuse opioids.

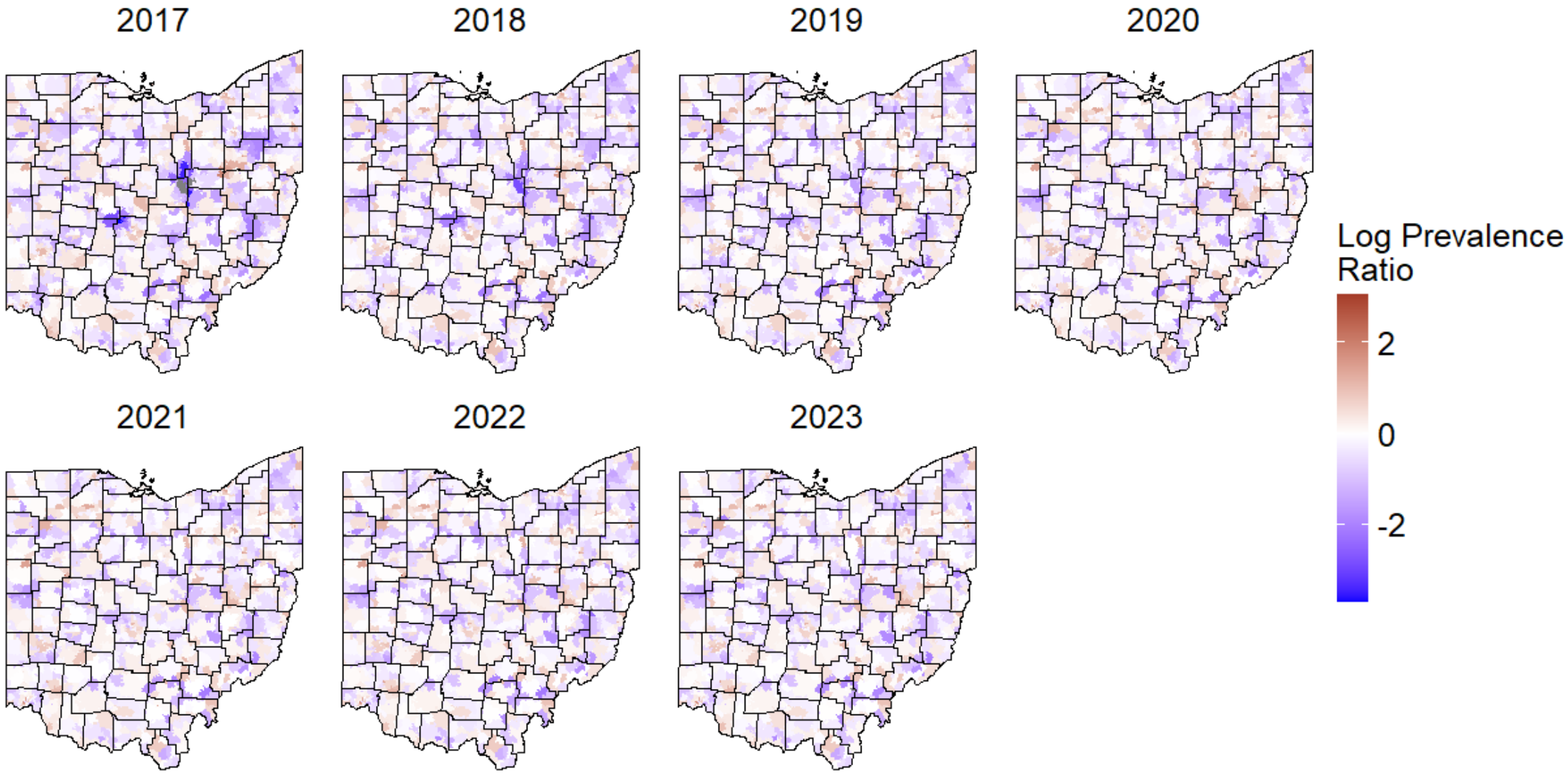


Figure 27: Log prevalence ratio of atom-level OMP relative to the county-level OMP.

### 5.3 State-Level Results

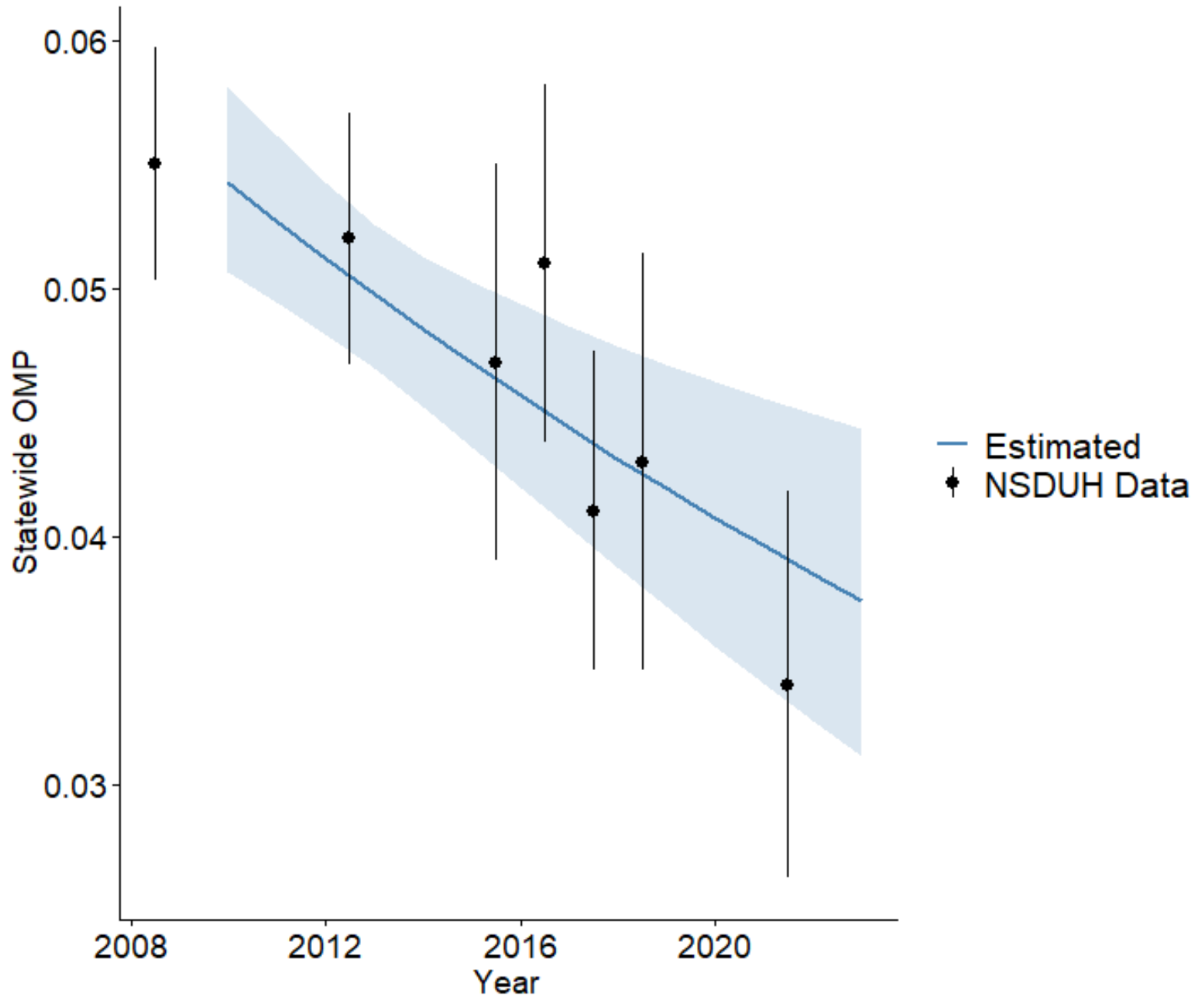


Figure 28: Estimated state-level OMP with the 95% credible interval shown as the shaded region. The NSDUH survey estimates are shown as points plotted at the midpoint of the time period of the survey. The error bars are 1.95 standard deviations above and below the estimate, using the standard deviations provided by NSDUH.



\end{document}